\documentclass[twocolumn,aps,prb,superscriptaddress,longbibliography]{revtex4-1}
\usepackage[latin9]{inputenc}
\usepackage{amsmath,tikz,graphicx}
\usepackage{esint}
\usepackage[colorinlistoftodos]{todonotes}
\makeatletter
\usepackage{H1}

\newcommand{\bra}[1]{ \langle{{#1}}|}
\renewcommand{\ket}[1]{|{{#1}}\rangle}

\makeatother

\begin{document}

\title{Entanglement growth after inhomogenous quenches}

\author{Tibor Rakovszky}

\affiliation{Department of Physics, T42, Technische Universit{\"a}t M{\"u}nchen, James-Franck-Stra{\ss}e 1, D-85748 Garching, Germany}

\author{C.W.~von~Keyserlingk}

\affiliation{University of Birmingham, School of Physics \& Astronomy, B15 2TT,
UK}

\author{Frank Pollmann}

\affiliation{Department of Physics, T42, Technische Universit{\"a}t M{\"u}nchen, James-Franck-Stra{\ss}e 1, D-85748 Garching, Germany}
\affiliation{Munich Center for Quantum Science and Technology (MCQST), Schellingstr. 4, D-80799 M\"unchen, Germany }

\begin{abstract}
We study the growth of entanglement in quantum systems with a conserved quantity exhibiting diffusive transport, focusing on how initial inhomogeneities are imprinted on the entropy. We propose a simple effective model, which generalizes the minimal cut picture of \citet{JonayNahum} in such a way that the `line tension' of the cut depends on the local entropy density. In the case of noisy dynamics, this is described by a Kardar-Parisi-Zhang (KPZ) equation coupled to a diffusing field. We investigate the resulting dynamics and find that initial inhomogeneities of the conserved charge give rise to features in the entanglement profile, whose width and height both grow in time as $\propto\sqrt{t}$. In particular, for a domain wall quench, diffusion restricts entanglement growth to be $S_\text{vN} \lesssim \sqrt{t}$. We find that for charge density wave initial states, these features in the entanglement profile are present even after the charge density has equilibrated. Our conclusions are supported by numerical results on random circuits and deterministic spin chains.

\end{abstract}

\maketitle

\section{Introduction}

Understanding how, and under what conditions, closed quantum systems approach thermal equilibrium due to their own unitary dynamics has been at the center of much recent attention, both theoretically and experimentally~\cite{Rigol2008,CalabreseCardy06,ETHreviewRigol16,GogolinReview,Kaufman794}. One important insight is to focus on the reduced density matrices of sufficiently small subsystems, which relax to a thermal Gibbs state with temperature, chemical potential, etc. set by the initial conditions, in systems where the Eigenstate Thermalization Hypothesis is satisfied~\cite{Deutsch91,Srednicki94,ETHreviewRigol16}. These long-time states have a von Neumann entropy that is extensive in subsystem size, in accordance with the prediction of thermodynamics. For a closed system, this entropy comes entirely from entanglement between the subsystem and its environment - in this sense we can say that a thermalizing quantum system acts `as its own bath'. Therefore an important part of understanding the mechanisms of thermalization is to describe how entanglement between subsystems builds up as a consequence of unitary dynamics, a question that has recently become partially amenable to experimental probes in systems of cold atoms, through the measurement of so-called R\'enyi entropies~\cite{DemlerEntanglement,ZollerEntanglement,Islam15,Kaufman794,Elben18}.

A simple picture of entanglement growth was developed for non-interacting and critical one-dimensional systems~\cite{CalabreseCardy05,CalabreseCardy07} and then generalized to generic quantum integrable systems~\cite{Alba2017_1,Alba2017_2}. In this picture, entanglement is carried by pairs of quasi-particles with opposite momenta that are present in the initial state, leading to linear growth of entanglement entropies for a contiguous subsystem with eventual saturation to a volume law. This description crucially relies on the notion of infinitely long-lived quasi-particle excitations. However, linear growth of entanglement is in fact more generic and has been shown to apply quite generally to so-called `global quenches'~\cite{CalabreseCardy06}, even in systems which do not admit a quasi-particle description, and exhibit diffusive transport~\cite{HyungwonHuse}. More recently, important advances have been made in explaining this linear growth from a a coarse-grained hydrodynamic or `minimal cut' picture~\cite{Nahum16,JonayNahum}. This picture is supported by calculations in random circuits~\cite{Nahum17,ZhouNahum}, holography~\cite{Mezei18},  and results in spin chains, both numerical~\cite{JonayNahum} and analytical~\cite{Bertini2018}. As an aside, however, we note that not all measures of entropy grow ballistically or are directly amenable to the minimal cut picture, even in ergodic systems; in particular a recent work by the present authors \cite{DiffusiveRenyi} argued that in systems with diffusive transport, such as the ones studied below, higher R\'enyi entropies grow diffusively after a global quench.

\begin{figure}
	\includegraphics[width=1.\columnwidth]{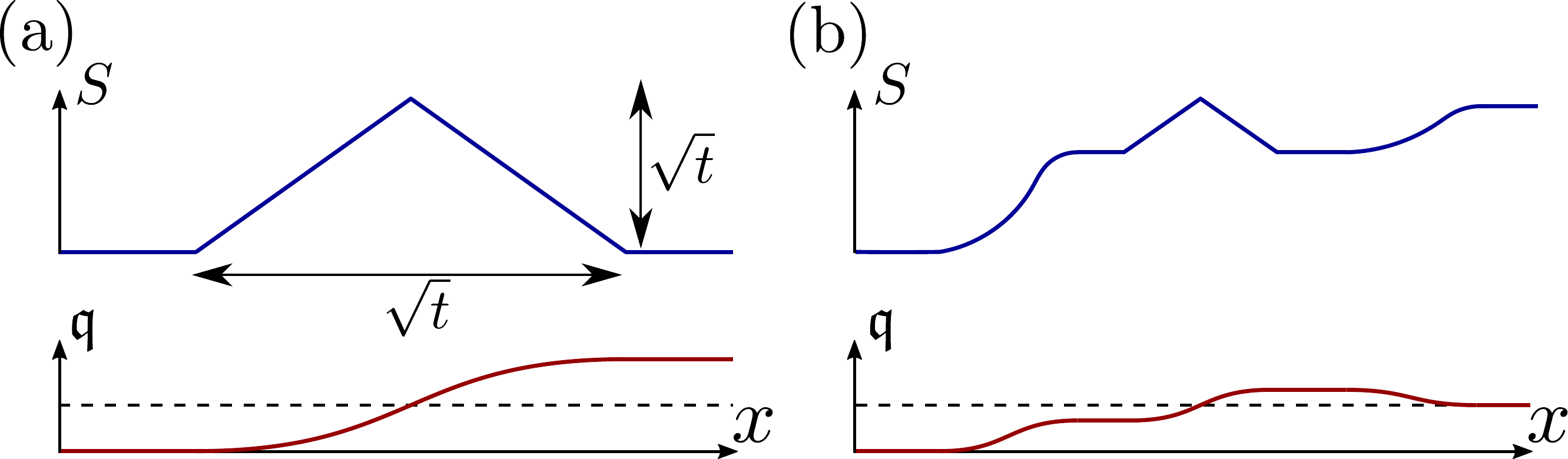}
	\caption{Sketch of the evolving charge density ($\mathfrak{q}$ and entanglement ($S$) profiles for inhomogenous initial states in a diffusive spin chain. The dashed horizontal lines denote half filling. (a) Starting from a maximally polarized domain wall state, entanglement builds up only in a region of size $\propto \sqrt{t}$ around the initial position of the domain wall, limiting the `height' of the entanglement profile to also be $O(\sqrt{t})$. (b) For a more generic initial state, with domains of different charge densities, the entanglement grows faster in regions where the charge density is closer to half filling, leading to an inhomogenous profile.} 
	\label{fig:sketch}
\end{figure}

Most of these results concern states that are invariant under spatial translations (hence the qualifier `global'). It is interesting to ask how the above picture of linearly growing entanglement entropy generalizes to the case of initial states with long-wavelength inhomogeneities. These are especially important if they involve changes in the densities of conserved quantities - these inhomogeneities than have to smooth out during time evolution through transport processes, implying that equilibrium can only be approached on time scales that are long compared to the associated wavelengths that characterize the initial state. An extreme case is that of a domain wall: an initial state where the density of one or more conserved quantity jumps between two extremal values in the middle of the system. Entanglement growth starting from such a domain wall was investigated for integrable systems, where both logarithmic~\cite{Eisler2009,Alba2014,Vidmar2017} and power law~\cite{Ljubotina2017,Karrasch2018} growth have been found for the entanglement. More recently, the quasi-particle picture has been extended to a variety of inhomogeneous initial states, using notions of generalized hydrodynamics, applicable to integrable systems~\cite{Bertini_2018,Bertini_2019,Mestyan_2019}. However, the case of generic interacting systems remains largely unexplored. 

The goal of the present paper is to investigate the question of entanglement production, starting from inhomogenous initial states in generic non-integrable systems, and elucidate how it is affected by the diffusive transport that is expected to be generic for the high temperature regime of non-integrable lattice systems~\cite{BLOEMBERGEN1949,DEGENNES1958,KADANOFF1963,Bohrdt16}. We approach this problem by combining the study of simple toy models with general considerations arising from hydrodynamics. In particular we consider local random unitary circuits with a U(1) symmetry, which have been used recently as a minimal model for dynamics in ergodic quantum many-body systems both with~\cite{OTOCDiff1,OTOCDiff2,DiffusiveRenyi} and without~\cite{Nahum17,RvK17} conserved quantities. Our approach here can be considered as a generalization of the hydrodynamic picture of Ref. \onlinecite{JonayNahum}, by considering how the entanglement couples to the simpler, diffusive hydrodynamic equations obeyed by the conserved densities such as particle number, energy etc. 

Using simple considerations, we conjecture an effective model that describes the coupling between entanglement growth and transport in our noisy random circuit setup, which can be interpreted as a Kardar-Parisi-Zhang (KPZ) equation~\cite{KPZ}, with a growth rate that is coupled to the diffusive variable. This implies that inhomogeneities in the original charge get imprinted into the growing `height' profile of the entanglement. In particular, we show that whenever there is a domain wall between two regions with different charge densities, the entanglement entropy near the interface has a component that grows in time as $\sqrt{t}$. The size of the region where this scaling applies also grows in time as $\sqrt{t}$ as the charge density smooths out due to diffusion. This is sketched in Fig.~\ref{fig:sketch}.  For a charge density wave initial state this process leads to an inhomogeneous entanglement profile across the system with a periodicity that is half the wavelength that characterizes the initial state. At longer times, as the charge distribution flattens, the entanglement profile also smooths out exponentially, but at a timescale that is parametrically larger than the time needed for the charge to equilibrate. We illustrate this general picture on a variety of different initial states, both for a random circuit model and for a deterministic system.

We also briefly discuss how the effects of diffusive charge transport show up in the von Neumann entropy even for states \emph{without} large scale inhomogeneities. In particular, we argue that the \emph{number entropy}, which is the component of the von Neumann entropy associated with the probability distribution of the conserved charge in a subsystem, has a characteristic growth, $\propto \log{t^{1/4}}$, which we associate to the diffusive spreading of correlations. This quantity is itself measurable in cold-atomic experiments~\cite{Lukin18}. 

The remainder of the paper is organized as follows. In Sec.~\ref{sec:setup} we introduce a charge-conserving local random circuit model that we use in the subsequent discussion. In Sec.~\ref{sec:nofluc} we analyze the effect of a single unitary gate to motivate a simple local update rule that couples entanglement growth with charge transport. This update rule is used to build an effective random surface growth model for entanglement dynamics, whose behavior we explore for a variety of inhomogeneous initial states. The results of this surface growth model are compared to numerics on both the random circuit model and on a deterministic spin chain in Sec.~\ref{sec:numerics}. In Sec.~\ref{sec:charge_entropy} we discuss the dynamics of the number entropy, which applies even in the case of initial states without long-range inhomogeneities. We conclude in Sec.~\ref{sec:discuss}. In App.~\ref{app:partfunc} we review the mapping of the calculation of the annealed average second R\'enyi entropy in the random circuit to a classical parition function and use it to arrive at a simple update rule in the absence of charge fluctuations. In App.~\ref{app:numerics} we show further numerical data, complementing the results of Sec.~\ref{sec:numerics}.


\section{Local random circuit with charge conservation}\label{sec:setup}

\begin{figure}
	\includegraphics[width=1.\columnwidth]{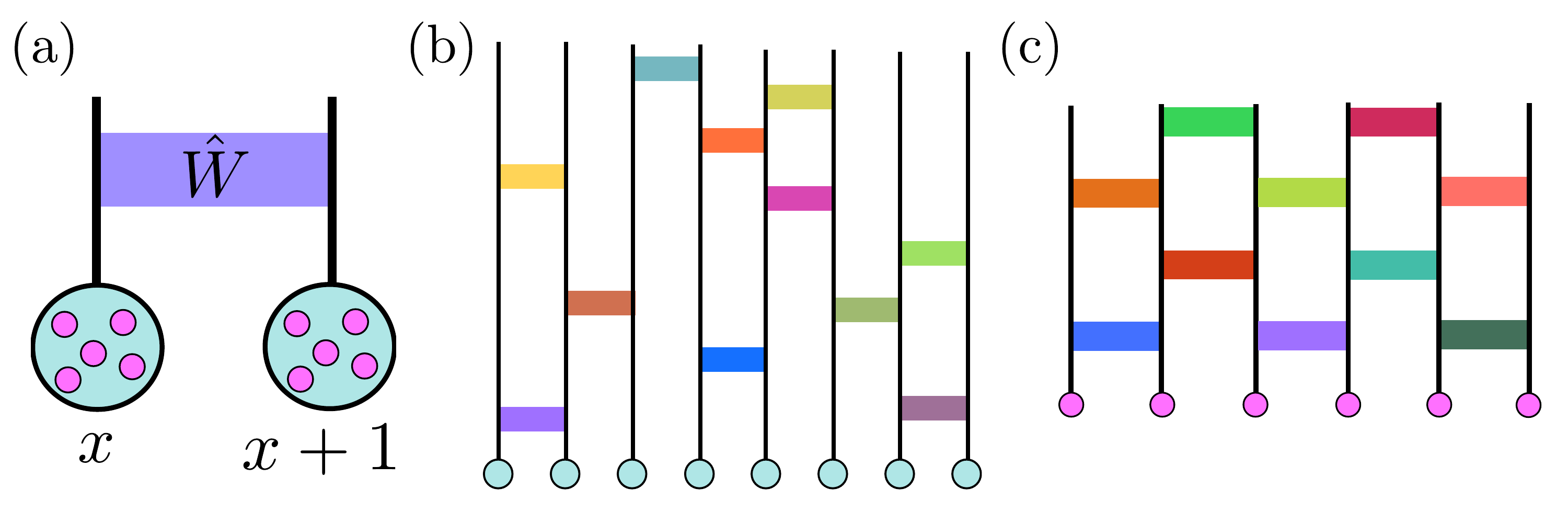}
\caption{Definition of the charge-conserving random circuit model. (a) A single unitary gate acts on two neighboring cells, each consisting of $N$ qubits, such that the total occupancy of the two cells is conserved. The full time evolution is generated by applying the gates (b) at randomly chosen bonds or (c) in a regular `brick wall' pattern.} 
	\label{fig:circuit_def}
\end{figure}

To investigate the relationship between entanglement growth and transport of conserved quantities, we turn to a simple minimal model that possesses all the main ingredients: a time evolution which is unitary, local in space, and has a conserved charge that obeys diffusive dynamics. This model was originally introduced in Ref. \onlinecite{OTOCDiff2}. Its basic building block is a collection of $N$ qubits (spin-$\frac{1}{2}$ degrees of freedom), which we refer to as a `cell' in analogy with the notion of a `fluid cell' in hydrodynamics or as a `site'. The physical system consists of $L$ such cells, arranged in a one-dimensional chain. We choose a specific basis for each qubit, and refer to these two basis states, $\ket{0}$ and $\ket{1}$, as `empty' and `filled' - the total number of filled sites is going to be our conserved quantity and we will refer to it as `charge'. On each cell, there are $\eta_a \equiv \binom{N}{a}$ states with charge $a=0,\ldots,N$. We evolve the system in discrete steps, by applying local unitary gates acting on pairs of neighboring cells in the chain, as indicated in Fig~\ref{fig:circuit_def}a. Each of these two-site gates is chosen such that it conserves the total charge on the two sites, i.e., it is block diagonal in the charge basis with a block of size $d_Q = \binom{2N}{Q}$ in the charge $Q$ sector. All blocks are independent and chosen Haar randomly. 

We consider two different circuit geometries. For the discussion in Sec.~\ref{sec:nofluc}, where we use the circuit to motivate a random surface growth picture of entanglement growth, we use a model where the two-site gates are applied on a randomly picked bond in each step, resulting in the irregular circuit shown in Fig.~\ref{fig:circuit_def}(b). Later on, in Sec.~\ref{sec:dw_N=1}, we will use a regular `brick wall' geometry, wherein odd numbered layers act on all the odd bonds of the chain while even numbered layers act on even bonds, as illustrated in Fig.~\ref{fig:circuit_def}(c). All the different gates are independently chosen from the charge conserving (i.e., block diagonal) random ensemble defined above. We denote averages over the different circuit realizations by $\overline{(\ldots)}$.

Such random circuit models have several advantages compared to usual Hamiltonian or Floquet systems, and have been used extensively to investigate a number of questions related to far-from-equilibium many-body systems~\cite{Nahum16,Nahum17,RvK17,Nahum18,ZhouNahum,OTOCDiff1,OTOCDiff2,ChanDeLuca1,ChanDeLuca2,ChanDeLuca3}. On the one hand they often admit exact analytical solutions, at least in certain limits (we will use the large-$N$ limit to simplify certain calculations). On the other hand, even when such exact treatment is not possible, by averaging over the randomness one can map the calculation of many relevant quantities to an effective `classical' model, which is often more tractable numerically than the original quantum problem. We will take advantage of such a mapping, which we outline in App.~\ref{app:partfunc}, in Sec.~\ref{sec:dw_N=1}. Last but not least, even when there is no direct computational advantage (e.g., in calculations of the number entropy in Sec.~\ref{sec:charge_entropy}), by averaging over random time evolutions, one can get access to the relevant universal behavior more easily then in any specific microscopic model - for example the random circuit shows diffusion of conserved quantities at all timescales.

\section{Surface growth model of entanglement growth}\label{sec:nofluc}

Our goal is to study the effect of an inhomogeneous charge distribution on the evolution of bipartite entanglement. We first discuss how the entanglement entropy $S(x)$ across the bond $x,x+1$ evolves under the application of a single charge-conserving unitary gate of the type introduced in the previous section. Based on these considerations, we conjecture a simple random surface growth model to capture certain universal features of entanglement growth. In Sec.~\ref{sec:numerics} we will compare the results of this surface growth model with numerics on various spin-$\frac{1}{2}$ chains. 

\subsection{Effect of a single gate}

We begin by considering the effect of a single charge-conserving 2-site gate, and use it to conjecture an effective update rule for the half-chain entanglement across the bond at these two sites. The main feature of this update rule, established using the subadditivity of entropy, is that the growth of entanglement is limited by the local entropy density, corresponding to the density of conserved charge. We also explore the possible interplay between charge density and the gradient of the half-chain entanglement.

We mainly focus on the \emph{half-chain von Neumann entropy}, $S_1(x,t) \equiv -\text{tr}\left(\rho \ln{\rho}\right)$, where $\rho$ is the reduced density matrix of the subsystem consisting of all sites $\leq x$, at time $t$. The von Neumann entropy is the $\alpha\to 1$ limit of the \emph{R\'enyi entropy}, $S_\alpha \equiv \frac{1}{1-\alpha}\ln{\text{tr}\left(\rho^\alpha\right)}$. We mostly focus on $S_1$ in this paper, which captures (the logarithm of) the typical eigenvalue of $\rho$, while later in this section we will briefly consider the case of $\alpha=0$, also known as the \emph{Hartley entropy}, which measures (the logarithm of) the rank of $\rho$. For this quantity, we are able to establish an exact update rule in the case when a charge-conserving gate is applied to a pair of neighboring sites with a fixed amount of total charge. Later on, in Sec.~\ref{sec:dw_N=1} we will also consider $S_2$, which is more amenable to numerical calculations in the random circuit setting.

\subsubsection{Von Neumann entropy $S_1$}

We begin by noting that the growth of $S_1$ is constrained by the local entropy density, $s_{x,t}$ (the von Neumann entropy of the subsystem consisting of the site $x$ alone), through the subadditivity condition~\cite{Nahum18},
\begin{equation}
 S_1(x,t+1)\leq S_1(x-1,t) + s_{x,t+1},
\end{equation}
where  we used the fact that the entanglement across the bond $x-1,x$ is unchanged by the 2-site gate acting on sites $x,x+1$ to replace $S_1(x-1,t+1)$ with $ S_1(x-1,t)$. The same argument holds when replacing $S_1(x-1,t)$ with $S_1(x+1,t)$ and $s_{x,t+1}$ with $s_{x+1,t+1}$, such that
\begin{align}\label{eq:subadbound}
    &S_1(x,t+1)  \nonumber \\
    &\leq \min{\big (}S_1(x-1,t)+s_{x,t+1},S_1(x+1,t)+s_{x+1,t+1}{\big )}.
\end{align}
What we will conjecture in the following is that replacing the inequality with an equality in Eq.~\eqref{eq:subadbound} provides a good qualitative description of the time evolution of the von Neumann entropy, particularly in situations where the charge has locally equilibrated. After local equilibration, we also expect that $s_{x,t+1}\approx s_{x+1,t+1} \approx(x,t)$, so that it can be pulled out of the $\min$ function. The resulting equation is already sufficient to capture the qualitative features we discuss below.

The bound in Eq.~\eqref{eq:subadbound} can be improved in certain circumstances, through the following line of reasoning, although there does not appear to be a sharp qualitative change in the physics in considering these subtler effects. In particular, we can consider a case where the two sites on which our unitary gate acts have a fixed total charge. This is motivated by the fact that if the charge distribution on the two sites in question is tightly peaked (as we would expect it to be at times in excess of the local equilibration time, in the large-$N$ limit), then the von Neumann entropy, which measures the size of typical Schmidt values, should be well approximated by focusing on Schmidt values corresponding to the average charge~\footnote{Note that this is in contrast with R\'enyi entropies $S_{\alpha > 1}$, which are strongly influenced by \emph{fluctuations} of the conserved charge, even for homogenous quenches, and can grow sub-ballistically as a consequence~\cite{DiffusiveRenyi}}
%

Thus, for the time being let us assume that the charge on sites $x,x+1$ takes a definite value $Q_{x,x+1} = Q$. In this case, the reduced density matrix on sites $\leq x$ is block diagonal in $Q_x$, the charge on site $x$. Let us denote the block with $Q_x=a$ as $p_a \rho_a$, where $p_a$ is the probability of having charge $a$ on site $x$, and $\text{tr}(\rho_a)=1$. Then the von Neumann entropy can be written~\cite{Lukin18}
\begin{equation*}
    S_1 = \sum_a p_a S_{1,a} -\sum_a p_a \ln{p_a},
\end{equation*}
where $S_{1,a} = -\text{tr}(\rho_a \ln{\rho_a})$ is the von Neumann entropy of the density matrix associated to charge $a$. Note that the last term is upper bounded by $\ln(Q+1)$, since $a$ takes values between $0$ and $Q$, and therefore should be strongly subleading, since the total von Neumann entropy will eventually increase to an extensive value. This means that to a good approximation, the entropy can be written as a weighted sum over entropies of different charge sectors. One can then apply subadditivity to each block separately
\begin{align*}
S&_{1,a}(x,t+1) \leq \\ &\leq  \min\left(S_{1,a}(x-1,t)+\ln{\eta_a},S_{1,a}(x+1,t)+\ln{\eta_{Q-a}}\right),
\end{align*}
where $\eta_a \equiv \binom{N}{a}$ is the dimension of the on-site Hilbert space with charge $a$. Here the entropies of the neighboring bonds also depend on $a$, since the projection to a given charge sector can in principle change the entanglement even away from the bond $x,x+1$. However, based on a simple numerical experiment involving a 4-site random MPS, we expect that if there is a gradient $S_1(x+1,t)-S_1(x-1,t)$ in the original state, then there should also be a a gradient $S_{1,a}(x+1,t)-S_{1,a}(x-1,t)$ in the projected state as well -- albeit one whose magnitude decreases with $a$. Therefore, one ends up with an upper bound that depends on a particular combination of the charge density and the entropy gradient. The same numerical experiment indicates that this bound is tighter then the bound in Eq.~\eqref{eq:subadbound}, at least away from half filling. This suggests that there is a possibility of a more complicated coupling between charge and entanglement, where the growth rate of the latter depends not only on the charge density, but also on the local gradient of $S_1(x,t)$. We will argue more rigorously for such a coupling in the case of $S_0$ below.  However, our numerical experiments suggest that the main qualitative features are already captured by the simpler bound in Eq.~\eqref{eq:subadbound}, ignoring these subtler effects. 

\subsubsection{Parameter counting argument for $S_0$}\label{sec:Hartley}

The above argument, regarding the coupling between charge and entanglement gradient, can be made more precise if we consider the Hartley entropy, $S_{\alpha=0}$, which equals the logarithm of the number of non-zero Schmidt values. Below we derive an update rule for this quantity under the effect of a generic unitary acting on a generic state with well defined charge on $x,x+1$. While the assumption of fixed local charge is not well motivated in the case of $S_0$, it is nevertheless useful to consider, as it sheds some light on the nature of the coupling between charge and entanglement gradient mentioned above.

Consider a matrix product state (MPS) with fixed 2-site charge $Q$ on sites $x,x+1$, and a bond dimension $\chi(x) \equiv e^{S_0(x)}$ across this bond. Note that $\chi(x)$ is nothing else but the number of non-zero Schmidt values for a decomposition between the two halves of the chain separated by this bond. One can then generalize the parameter counting argument of Ref. \onlinecite{Nahum16} to estimate the bond dimension $\tilde\chi(x)$ across the same bond after a charge-conserving unitary has been applied to these two sites~\footnote{While this parameter counting argument is not rigorous, in the sense that fine-tuned states or unitaries could violate it, it is expected to hold for generic states, which we indeed find numerically for random MPS.}. Since $Q$ is conserved, we can label the Schmidt values at $x,x+1$ (after applying the unitary) according to the amount of charge on site $x$. Let us denote this charge by $Q_x=a$. We will estimate the number of non-zero Schmidt values $\tilde \chi_a(x)$ in each $a$ sector separately, by equating the number of parameters in the new MPS tensors on sites $x,x+1$ with the number of equations defined by equating the new MPS to the time evolved one. 

We want to find new MPS tensors on both $x$ and $x+1$ that describe the state after it has been evolved with the unitary gate on these sites. For a given $a$, the two MPS tensors have $\chi(x-1)\tilde\chi_a(x) \eta_{a} + \tilde\chi_a(x) \chi(x+1) \eta_{Q-a} - \tilde\chi_a(x)^2$ parameters in total, where $\chi(x\pm1)$ are the bond dimensions on neighboring bonds and the last term takes into account the gauge freedom in choosing the MPS tensors. The number of parameters has to be sufficiently large to satisfy all the $\chi(x-1)\chi(x+1) \eta_{a} \eta_{Q-a}$ equations, coming from equating the new tensors with the original MPS evolved by a single 2-site gate. Equating the number of equations with the number of parameters gives a quadratic equation for $\tilde \chi_a(x)$. Taking the smaller of the two solutions and adding up the Schmidt values for different $a$ sectors gives the total bond dimension across a bond $x,x+1$ with charge $Q$, after applying the gate:
\begin{equation}
    \chi(x,t+1) = \sum_a \min\left(\chi(x-1,t)\,\eta_a,\chi(x+1,t)\,\eta_{Q-a}\right).
\end{equation}
i.e., we end up with the maximal bond dimension allowed by subadditivity, in each charge sector separately.

Without loss of generality, let $\chi(x+1) = \chi(x-1) e^{2\Delta}$, for some entropy gradient $\Delta \geq 0$ going across the bond (note that $e^{2\Delta} \leq d_Q$, due to subadditivity of entanglement). Then, we get the update rule for the Hartley entropy, $S_0(x,t) \equiv \log{\chi(x,t)}$, as
\begin{equation}\label{eq:param_counting}
    S_0(x,t+1) = \min\left(S_0(x-1,t),S_0(x+1,t)\right) + f_N(Q,\Delta),
\end{equation}
where $f_N(Q,\Delta) \equiv \log\left(\sum_{a=0}^{Q} \min(\eta_a,\eta_{Q-a} e^{2\Delta})\right)$, and we measure logarithms base $2^N$. Therefore we find a result that has more structure then the simple subadditivity bound in Eq.~\eqref{eq:subadbound}: the entropy increase depends not only on the local entropy, but also on the spatial derivative of $S_0(x,t)$ across the bond. This arises due to the fact that after fixing $Q_{x,x+1}$, we can count the entropy for each charge sector separately: having a gradient $\Delta$ across the bond allows for putting more charge on one site in order to increase the total entropy.

\begin{figure}[t!]
	\includegraphics[width=0.7\columnwidth]{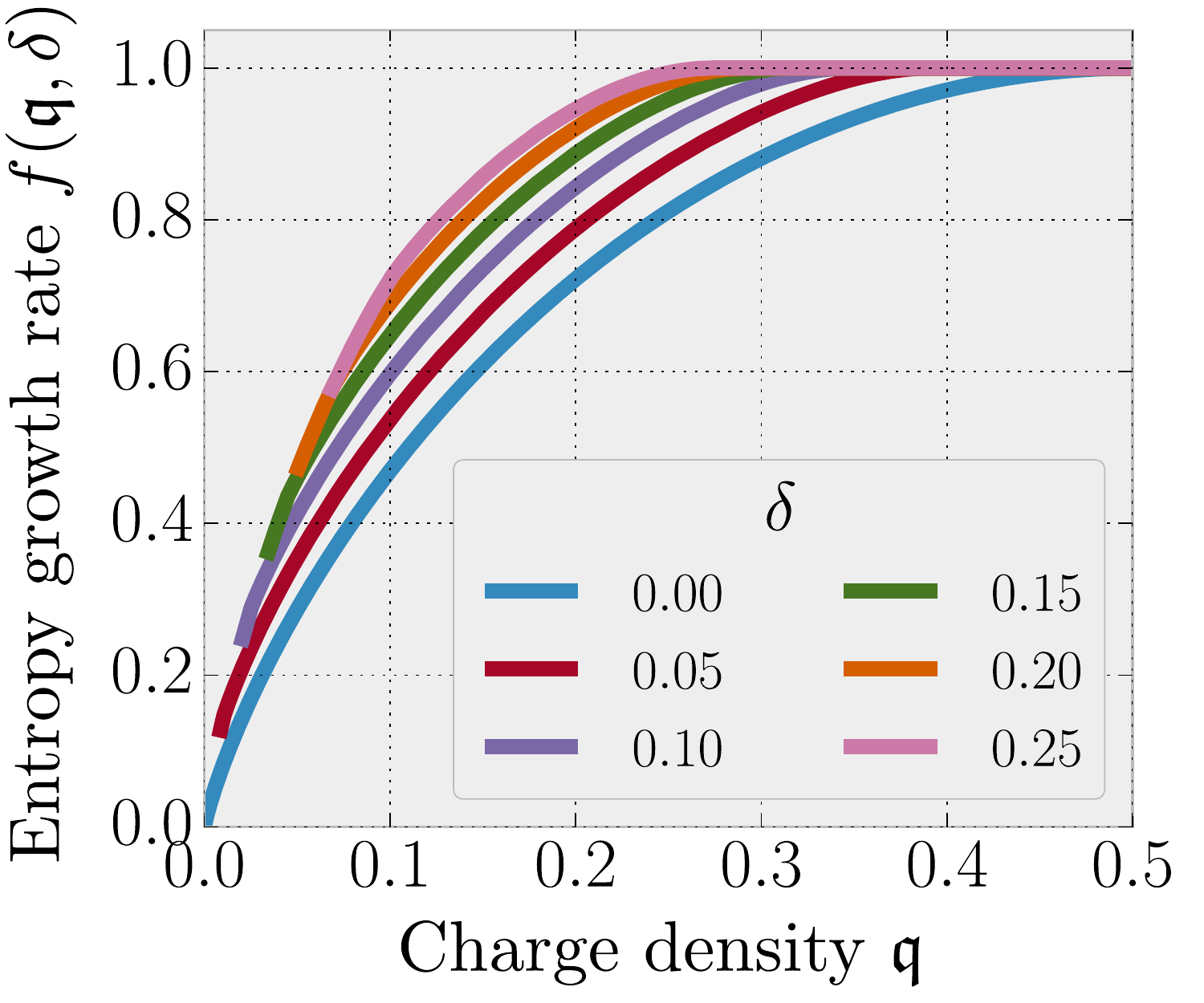}
	\caption{The entropy growth function $f(\mathfrak{q},\delta)$, appearing in the RHS of Eq.~\eqref{eq:param_counting}, in $N\to\infty$ limit, using rescaled variables $\mathfrak{q}\equiv Q/(2N)$ and $\delta = \Delta S/N$.} 
	\label{fig:Hartley}
\end{figure}

To get a sense of the behavior of the term $f_N(Q,\Delta)$ in Eq.~\eqref{eq:param_counting}, we take its large-$N$ limit. In this case $\eta_a = {N \choose a} \to e^{N h_2(a/N)}$, using the binary entropy function, $h_2(\xi) \equiv -\xi\ln{\xi}-(1-\xi)\ln(1-\xi)$. Defining rescaled variables, $\xi \equiv a/N$, $\mathfrak{q}\equiv Q/2N$ and $\delta = \Delta/N$, the condition $\eta_a = \eta_{Q-a} e^{2\Delta}$ translates to $h_2(\xi) = h_2(2\mathfrak{q}-\xi)+2\delta$. Let $\xi_*(\mathfrak{q},\delta)$ denote the solution to this equation when it exists, and $\xi_* = 2\mathfrak{q}$ otherwise. The sum over $a$ in the definition of $f_N$ can then be replaced by a pair of integrals over $\xi$, on intervals $[0,\xi_*]$ and $[\xi_*,2\mathfrak{q}]$. In the limit $N\to\infty$ these integrals can be approximated by the maximal value iside the interval, giving the result 
\begin{equation}\label{eq:param_count_largeN}
    f(\mathfrak{q},\delta) \equiv f_{N\to\infty}(2N\mathfrak{q},N\delta) = 
    \begin{cases}
    \frac{h_2(\xi_*)}{\ln{2}} &\text{ if } \xi_* \leq 1/2 \\ 
    1 &\text{ if } \xi_* > 1/2.
    \end{cases}
\end{equation}
Fig.~\ref{fig:Hartley} shows the resulting function $f(\mathfrak{q},\delta)$ as a function of the charge density $\mathfrak{q}$ for a variety of $\delta$. Its most notable feature is that for $\delta > 0$ it becomes constant for sufficiently large values of $\mathfrak{q}$.

The arguments above, in particular the result~\eqref{eq:param_counting}, indicate a complicated coupling between charge and entanglement entropy, where the local growth rate of the bipartite entanglement depends on a particular combination of local charge density and the entropy gradient. However, in the physical situations we consider in the next section, we find that this dependence on the entropy gradient does not change the qualitative features of entanglement growth, and in practice one can replace the constant term in the RHS with just the on-site entropy $h_2(\mathfrak{q})$, which is the original subadditivity bound in Eq.~\eqref{eq:subadbound}. As we argue below, in Sec.~\ref{sec:inhom}, this can be understood from the fact that in a coarse-grained continuum description the difference only shows up in the form of higher order subleading corrections. Nevertheless, it would be interesting to understand whether there are physical situations where such a coupling can still play an important role.

\subsection{Surface growth model}\label{sec:inhom}

The equation of motion in Eq.~\eqref{eq:param_counting} is a direct generalization of the local update rule for a non-symmetric random gate, derived in Ref. \onlinecite{Nahum16}; the only difference being that the constant growth term on the RHS now depends on space and time, through the local charge, and also on the entropy gradient. We can then construct an effective, `surface growth' model of entanglement growth, along the same lines as in Ref. \onlinecite{Nahum16}: in each time step we apply the update rule on a randomly chosen bond, using the function $f(\mathfrak{q},\delta)$ derived in Eq.~\eqref{eq:param_count_largeN}, to update the half-chain entanglement $S(x,t)$ (the `height'). At the same time we update the local charge densities $\mathfrak{q}(x) \equiv Q_x / N$ as $\mathfrak{q}(x),\mathfrak{q}(x+1)\to(\mathfrak{q}(x)+\mathfrak{q}(x+1))/2$. This leads to a coupled stochastic evolution between $\mathfrak{q}(x)$ and the entropy $S(x,t)$. Below, we investigate the behavior of this stochastic model, which we compare to simulation of quantum systems in Sec.~\ref{sec:numerics}.

\subsubsection{General considerations}

As stated before, we find numerically that the coupling between the charge and the entropy gradient does not affect the main qualitative features of the evolution, therefore in practice one can replace $f(\mathfrak{q},\delta)$ with $f(\mathfrak{q},0) = h_2(\mathfrak{q})$, the local binary entropy associated to the charge density. This leads to a simplified update rule
\begin{equation}\label{eq:large_N_eom}
    S(x,t+1) = \min\left(S(x-1,t),S(x+1,t)\right) + h_2(\mathfrak{q}).
\end{equation}
In the continuum limit this corresponds to a KPZ equation for the entanglement (as in Ref. \onlinecite{Nahum16}), but one that is coupled to the diffusion equation for the charge as
\begin{align}\label{eq:kpz_diff}
\partial_t S(x,t) &= \nu \partial_x^2 S - \frac{\lambda}{2} (\partial_x S)^2 + s(x,t) (c + \zeta(x,t)); \nonumber\\
\partial_t \mathfrak{q}(x,t) &= D \partial_x^2 \mathfrak{q} + \zeta_\mathfrak{q}(x,t)
\end{align}
where $s(x,t)\equiv h_2(\mathfrak{q}(x,t))$ is the entropy density, $\zeta(x,t)$ is uncorrelated white noise, $\zeta_\mathfrak{q}(x,t)$ is a noise field consistent with charge conservation (model B dynamics~\cite{Tauber2007}) and the diffusion constant $D$ is $1/2$ for the random circuit model~\cite{OTOCDiff1,OTOCDiff2}. Note that is we kept the function $f(\mathfrak{q},\delta)$ in Eq.~\eqref{eq:large_N_eom} (instead of replacing it with $f(\mathfrak{q},0) = h_2(\mathfrak{q})$), the highest order terms we would need to add to the continuum description would be of the form $(\partial_x \mathfrak{q})^2 (\partial_x S)^2$, which are subleading in the long wavelength limit, therefore we are justified in dropping them.

The above random surface growth model also admits an interpretation as a directed polymer problem, in the spirit of Ref. \onlinecite{JonayNahum}. As shown there, on the longest length and time scales, the entanglement resulting from Eq.~\eqref{eq:large_N_eom} can be rewritten as the energy of a minimal energy polymer, characterized by an `entanglement line tension' $\varepsilon(v)$, with $v$ being the slope of the polymer, representing a space-tim cut through the unitary circuit. Since our update rule differs from the one in Ref. \onlinecite{Nahum16} by having a constant term $s(x,t)$, rather then $1$, on the RHS, one has to rescale $S \to S/s$ to get the same physics locally. This implies that the line tension gets rescaled as $\varepsilon(v) \to s(x,t)\varepsilon(v)$. Therefore the calculation of entanglement growth becomes that of finding a polymer with minimal energy in a space-time dependent background, where the background itself contains a deterministic evolving part, governed by the diffusion equation, as well as random noise. In particular, parts of the systems with very low/high fillings act as bottlenecks for the entanglement growth of nearby regions~\footnote{This is similar to the mechanism proposed for entanglement growth in disordered Griffith phases where the bottleneck is provided by localized regions which act as `weak links', see Ref. \onlinecite{Nahum18}}. On the other hand, the charge density itself undergoes diffusion, tending toward a more homogenous distribution. The entropy density follows the charge distribution, leading to a speed-up of entanglement growth in regions where it gets closer to half filling and vice versa. 

We expect this generalized minimal cut picture to apply also to systems without noise, much like its original version which did not take conservation laws into account~\cite{JonayNahum}. In our case this would correspond to an equation of motion analogous to Eq.~\eqref{eq:kpz_diff}, but with the noise terms omitted. The original minimal cut picture can also be generalized to higher spatial dimensions, replacing the polymer with a `membrane'~\cite{Nahum16,JonayNahum,Mezei18}. We expect that the model we consider here similarly generalizes, with a minimal membrane whose local surface tension depends on the entropy density at a given position and time.

\subsubsection{Application to various initial states}

To see the effect that the diffusion of charge has on entanglement growth, let us first consider the paradigmatic example of a maximally polarized domain wall initial state, wherein all sites on the left half of the chain are empty ($\mathfrak{q}(x\leq 0,t=0) = 0$) and all sites on the right half are filled ($\mathfrak{q}(x>0,t=0) = 1$). We imagine working in the thermodynamic limit, so that $x\in\mathbb{Z}$ and we can ignore boundary effects. At long times and large distances we expect the charge profile $\mathfrak{q}(x,t)$ to be well approximated by the solution of the continuum diffusion equation, which for these initial conditions reads $\mathfrak{q}(x,t) = \left( 1 + \text{erf}(x/\sqrt{Dt}) \right) / 2$, where $\text{erf}(x)$ is the error function~\footnote{This is most easily seen by taking the spatial derivative of the diffusion equation, which results in the same equation for $\partial_x \mathfrak{q}$. The initial condition for $\partial_x \mathfrak{q}$ is given by a delta function, $\partial_x \mathfrak{q}(x,0) = \delta(x)$, for the domain wall, which becomes a Gaussian of width $\sqrt{Dt}$ at time $t$. Integrating it up gives the result for $\mathfrak{q}(x,t)$ stated in the text.}. The important property of this solution is that the width of the domain wall scales diffusively, as $\propto \sqrt{t}$, with time, and consequently the local entropy density $s(x,t)$ is also a function solely of $x/\sqrt{Dt}$. This indicates that significant entanglement growth can only occur within a region of order $\mathcal{O}(\sqrt{Dt})$ around the origin. At long times this diffusive growth is very slow compared to the linear growth of the entanglement surface that one would get for a homogeneous charge distribution. This means that the charge diffusion acts as a bottleneck for the entanglement growth: for the entanglement in the middle to grow further it has to `wait' for the width of domain wall to increase. As a consequence, the height itself grows as $S(0,t) \propto \sqrt{t}$, as sketched in Fig.~\ref{fig:sketch}(a). 

\begin{figure}[t!]
	\includegraphics[width=1.\columnwidth]{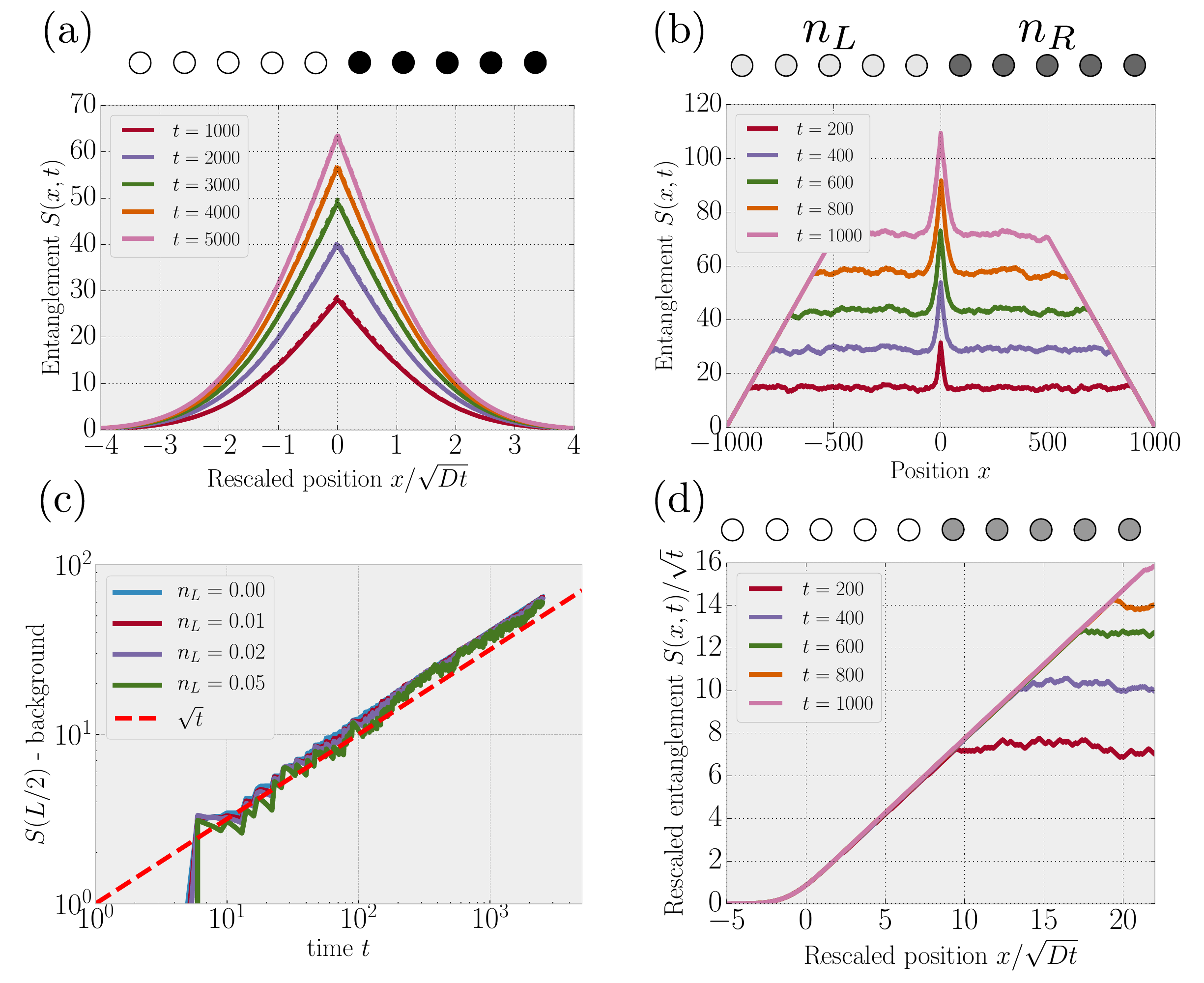}
	\caption{Entanglement growth generated by applying the update rule~\eqref{eq:large_N_eom} to a randomly chosen bond at each step. (a) For an initial domain wall state entanglement grows as $\propto\sqrt{t}$ and is close to the maximum value allowed by subadditivity. As a consequence, random fluctuations are strongly suppressed. (b) For a generalized domain wall, with charge densities $n_L$ and $n_R = 1-n_L$ on the two sides, rhe `bump' appears on top of a linearly growing background where KPZ fluctuations are present. (c) Subtracting the average height of this background, we observe that the size of the bump itself always grows $\propto \sqrt{t}$. (d) For an asymmetric domain wall, $n_L = 0$, $n_R = 1/2$, there is a region with $\propto\sqrt{t}$ entanglement interpolating between the two bulk regions.}
	\label{fig:dw_large_N}
\end{figure}

By performing the stochastic surface growth numerically, we find a scaling collapse of the form $S(x,t) = \sqrt{Dt}\,f(x/\sqrt{Dt})$ for some function $f$, as shown in Fig.~\ref{fig:dw_large_N}(a). At long enough times, $Dt \gg 1$ (in units of the lattice spacing), the surface profile is always close to being saturated to the maximum value allowed by the charge distribution, with a distance between the two becoming constant. This also means that there is no allowed `space' for the surface to develop large random fluctuations, which would be present for a flat surface - such KPZ fluctuations are suppressed by the slow dynamics of the charge that is coupled to the surface growth. In the opposite limit of $Dt \ll 1$ (which we cannot test in the random circuit model), the growth of the domain wall appears fast compared to the speed of the surface growth: the entanglement profile near the origin is not yet sensitive to finite width of the domain wall and we expect it to grow the same way as it would for a homogenous state at half filling.

While the domain wall constitutes  an extreme case, where the charge densities at the two infinities take the two extreme values, $0$ and $1$, a similar behavior occurs more generally if we take an initial state with a jump in the average charge at the origin. The simplest generalization is to take an initial state where $\mathfrak{q}(x\leq 0) = n$ and $\mathfrak{q}(x > 0) = 1-n$ for some $0 < n < 1/2$. In this case the dynamics of entanglement far from the origin, at positions $x \gg \sqrt{Dt}$ is the same as for a homogenous state with charge density $n$, i.e., it grows linearly with the rate set by $n$ and exhibits KPZ fluctuations. However, in a diffusively growing region around the origin, where the charge density has started equalizing, the entanglement grows faster and exhibits a `bump' on top of the linearly growing background. Both the spatial size and the height of this bump is once again increasing as $\sqrt{t}$ (see Fig.~\ref{fig:dw_large_N}(c)) and the KPZ fluctuations are suppressed in this region. This is shown in Fig.~\ref{fig:dw_large_N}(b). 

\begin{figure}[t!]
	\includegraphics[width=1.\columnwidth]{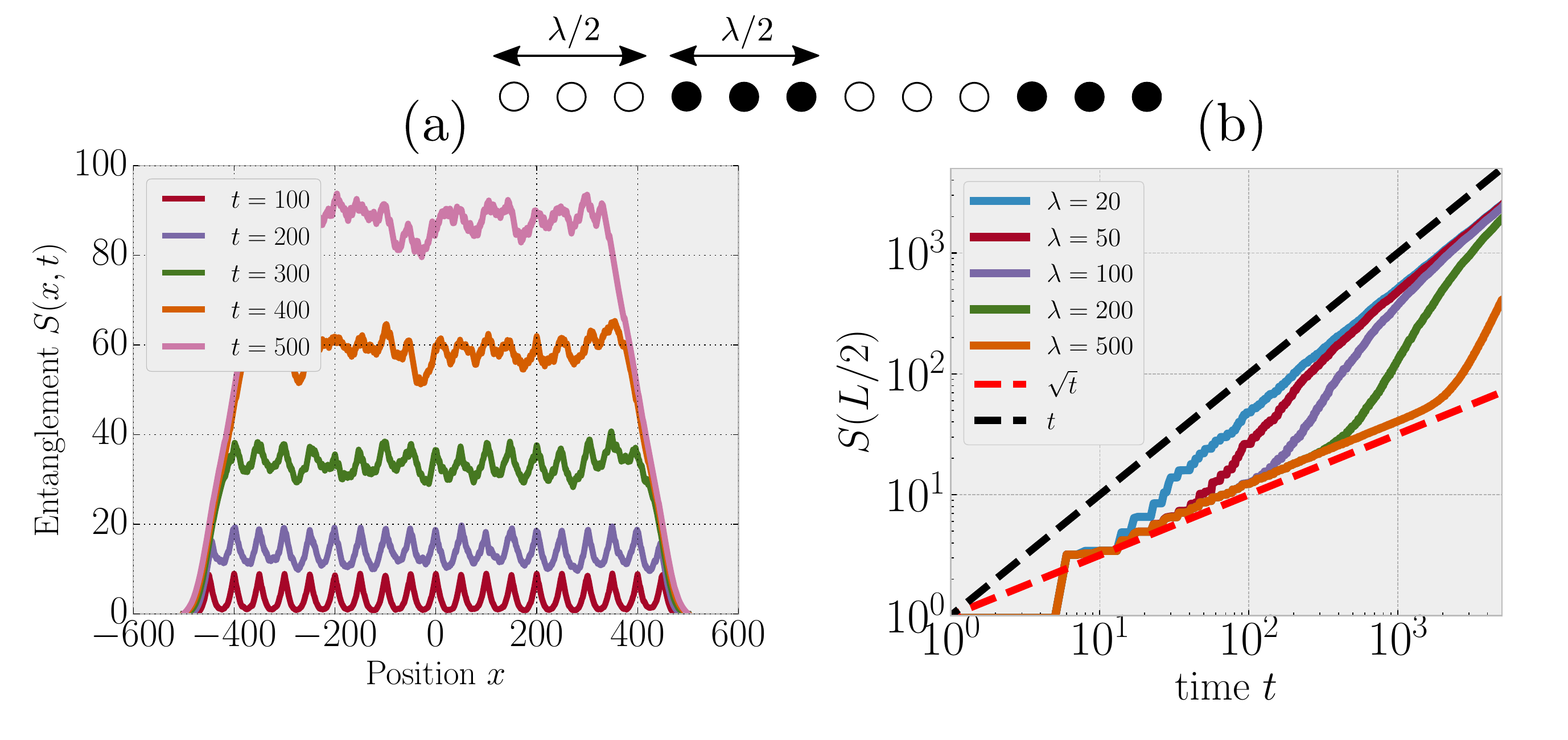}
	\caption{Entanglement growth from the random surface growth model with charge density wave initial state of equally spaced domain walls at a distance $\lambda/2 = 50$ sites. (a) the initial melting of the domain walls creates an inhomogenous entanglement profile. This subsequently smooths out once the charge has equilibrated. At the same time KPZ behavior takes over, with fluctuations growing in time. (b) Consequently, the overall growth rate at a domain wall has a cross-over from $\sqrt{t}$ to $t$ at a time $\propto \lambda^2$ when the domain walls melt into each other (shown for $L=5000)$.}
	\label{fig:cdw_large_N}
\end{figure}

Another, slightly different scenario where the diffusion plays an important role, is an interface between two regions of different entropy densities. An extreme example of this is when a region at half filling, such as a N\'eel state, expands into empty space. In this case, in the middle of the half-filled region, far from the interface, we once again get linear growth with the usual KPZ fluctuations. Far on the other side of the interface the charge density vanishes and so does the entanglement. Near the interface there is a diffusively growing region that interpolates between these two extremes, where the entanglement obeys the scaling $S(x,t) = \sqrt{Dt}\,\,g(x/\sqrt{Dt})$, for some scaling function $g$. The middle region where this scaling is valid then penetrates linearly into the half-filled region. This is shown in Fig.~\ref{fig:dw_large_N}(d). A similar situation would occur if the average charge densities on the two halves of the chain are initially $0 < n_L < n_R < 1-n_L$. In this case there is linear growth on both sides, but with different growth rates, so that there are two plateaus of different heights with an interpolating region in-between.

\begin{figure}[t!]
    \includegraphics[width=1.\columnwidth]{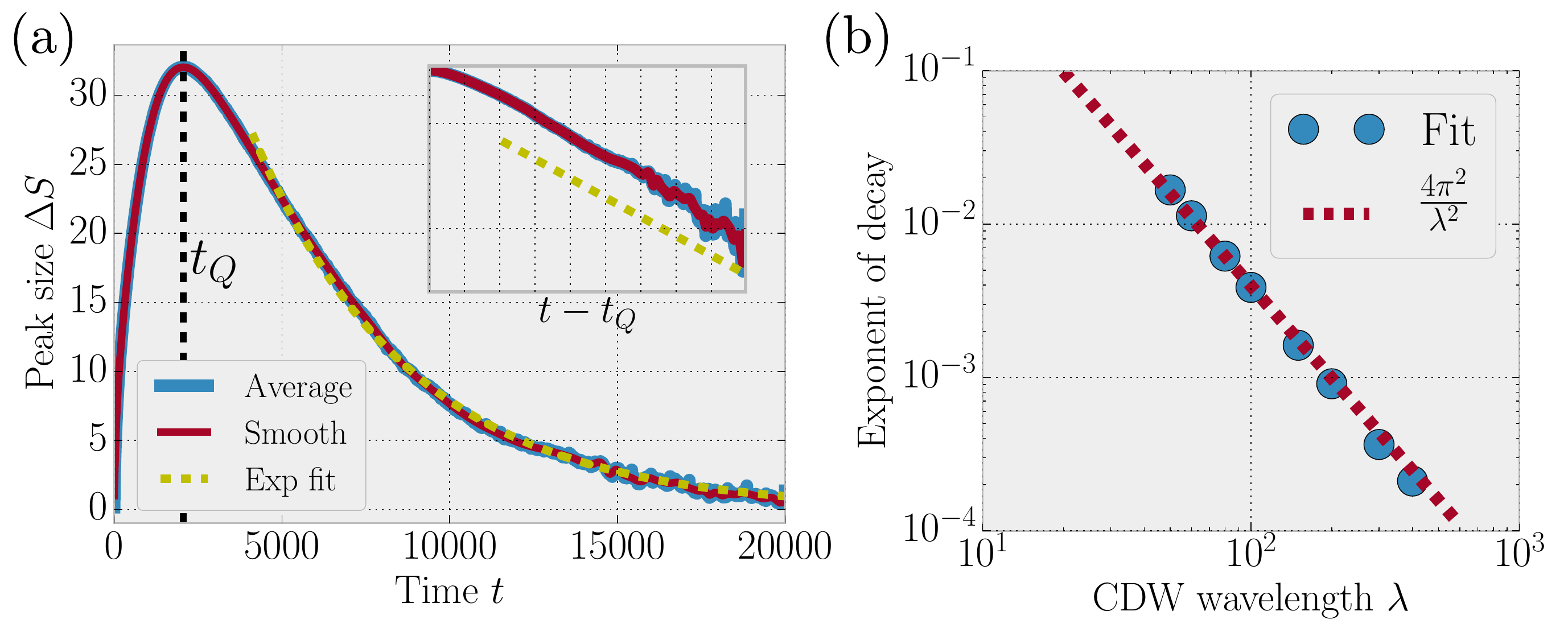}
	\caption{Smoothing out of the inhomogeneities of $S(x,t)$, starting from a charge density wave initial state. We calculate the size of the peaks due to the initial inhomogenous growth (see Fig.~\ref{fig:dw_large_N}(b)), averaged over all peaks in a system and over $10$ realizations of the stochastic dynamics. a) We find that the size of the peaks grows up to some timescale $t_Q$ (denoted by the dashed vertical line) when the different domains melt together and the profile starts smoothing out exponentially. Data is shown for a chain with $L=40000$ sites and CDW wavelength $\lambda = 400$. The thick blue curve is numerical data for the average peak size, the thin red curve is a numerically smoothened extrapolating curve and the dashed yellow line is an exponential fit. The inset shows the same data as a function of $t-t_Q$. b) Exponents from the exponential fit for different wavelengths $\lambda$. We find that a decay $\propto e^{-\frac{4\pi^2}{\lambda^2}t}$.} 
	\label{fig:smooth_large_N}
\end{figure}

The initial states described above all have charge-imbalances between the left and right halves of the chain, such that in the thermodynamic limit it would take an infinite amount of time for them to become homogeneous. We can also consider initial states which have charge-inhomogeneities on some parametrically smaller length scale $1 \ll \lambda \ll L$. An example of such a state is a step-like charge density wave (CDW) consisting of domains of $\lambda/2$ filled sites followed by $\lambda/2$ empty ones in a regular pattern. At short times, $t \ll \lambda^2 / 4D$, each domain wall in the initial state evolves independently, in the way described above, while far from the domain walls there is no dynamics. This results in an entanglement profile with peaks separated by a distance $\lambda/2$ from one another, whose width and height grows as $\sqrt{t}$. Once the peaks start to overlap, at times $t_Q \sim \mathcal{O}(\lambda^2 / 4D)$, i.e., when the domain walls start to melt together, the bottleneck disappears and the entanglement growth speeds up from $\sqrt{t}$ to linear in $t$ (see Fig.~\ref{fig:cdw_large_N}(b)). At the same time the entanglement profile starts to smooth out: the difference between the top and the bottom of a peak, $\Delta S \equiv S(0,t) - S(\lambda/4,t)$, decreases to zero, as shown by Fig.~\ref{fig:cdw_large_N}. We find numerically (see Fig.~\ref{fig:smooth_large_N}) that the average size of the peaks in the entanglement profile decreases exponentially, as $\Delta S \propto e^{-\frac{4\pi^2}{\lambda^2}t}$. We associate this with the first (diffusive) term on the right hand side of the first equation in Eq.~\eqref{eq:kpz_diff}.

Using the results above we can associate two distinct time scales to the evolution of the entanglement starting from a CDW, made out of regions of different charge densities of typical size $\lambda$. First there is the aforementioned time scale $t_Q \sim \lambda^2 / D$, until which each domain wall evolves independently. By this time the entanglement profile develops peaks of size $\Delta S \propto \sqrt{Dt_Q} \propto \lambda$. At times $t\approx t_Q$ a second time regime begins, wherein the entanglement features smooth out exponentially as $\Delta S \propto \lambda e^{-8\pi^2\frac{t-t_Q}{t_Q}}$. The time needed for the entanglement profile to become sufficiently smooth, i.e., $\Delta S \ll 1$ is therefore $\frac{t-t_Q}{t_Q} \gg \log{\lambda}$, or equivalently $t \gg \frac{\lambda^2}{D} (1 + \log{\lambda})$. Note that this is parametrically larger than the time scale it takes for the charge-distribution to become approximately flat, which merely requires $\frac{t-t_Q}{t_Q} \gg 1$.

To summarize our findings in this section, let us consider a rather broad set of initial states, that consists of different domains of typical size $\lambda$, each domain roughly homogeneous with some specific average charge density. For such a state, we can distinguish four different time scales. I) There is some short initial time scale $t_\text{loc}$ associated to reaching local equilibrium. II) This is followed by the transport of charge from domains of high density to neighboring regions of lower density. In the middle of the domains the entanglement grows linearly at the rate set by the local entropy density of the region. At the interfaces of different domains the transport of charge is accompanied by the appearance of `bumps' in the entanglement profile. These bumps grow as $\sqrt{t}$ with respect to the entanglement in the middle of the domains which they separate (the one with the smaller entropy density if those are different). III) This bumpy entanglement profile develops up to times $t_Q \sim \mathcal{O}(\lambda^2/D)$, at which point the domains start to melt completely into one another. The entanglement growth at the domain walls speeds up to linear in time and the entanglement profile smooths out. There is another timescale associated to this smoothing which scales as $t_Q(1+\log{\lambda})$. IV) Eventually there is a fourth time regime, when the entanglement stops growing and saturates to its equilibrium value (assuming the system is finite). Note that the III) regime is only present if there is a separation of scales $\lambda \ll L$. If the initial inhomogeneities are on the scale of the entire system then equilibration of the charge density and the saturation of the entanglement profile happens simultaneously. 

\section{Spin-$\frac{1}{2}$ chains}\label{sec:numerics}

We now compare the predictions of the simple effective model of the previous section with results on spin-$\frac{1}{2}$ chains. We first take the a version of the $N=1$ random circuit model defined in Sec.~\ref{sec:setup}, one where the arrangement of random gates follows a regular brick-wall pattern (Fig.~\ref{fig:circuit_def}(c)). In this case we can compute efficiently the annealed average of the $2^{\text{nd}}$ R\'enyi entropy (defined below). While R\'enyi entropies are expected to have very different dynamics from the von Neumann entropy in the case of a \emph{global} quench~\cite{DiffusiveRenyi}, we find that the effects associated to charge inhomogeneities are similar to those predicted by our surface growth model for the latter. In particular, for a domain wall initial state we find a scaling collapse of the entanglement profile of the form $S(x,t) = \sqrt{t} f(x/\sqrt{t})$. In Sec.~\ref{sec:floquet_def} we compare to exact results on a non-random, periodically driven spin chain.

\subsection{Spin-$\frac{1}{2}$ circuit model}\label{sec:dw_N=1}

The simplified model of the previous section relied on ignoring local fluctuations of the charge, and using only the average local density as the only relevant variable. Here we show that the features associated with inhomogeneities are quite similar even if we consider the spin-$\frac{1}{2}$ random circuit where fluctuations in local charge density are expected to be even larger. We take now a circuit where the unitary gates are applied in the regular `brick wall' pattern shown in Fig.~\ref{fig:circuit_def}(c), and each site contains a single conserved spin ($N=1$). 

We consider the \emph{annealed average} of the second R\'enyi entropy, $S_2^{(a)} \equiv -\ln\left(\overline{\text{tr}(\rho_A^2)}\right)$. The calculation of this quantity can be mapped the problem of evaluating a classical partition function~\cite{DiffusiveRenyi}. For $N=1$ this classical partition function can be evaluated by representing it as a two-dimensional tensor network, as we review in App.~\ref{app:partfunc}. This allows us to treat the dynamics of $S_2^{(a)}$ for much larger systems and longer times then those available in the original quantum problem. The annealed average R\'enyi entropy provides a lower bound on the von Neumann entropy, $S_2^{(a)}(t) \leq \overline{S_2(t)} \leq \overline{S_\text{vN}(t)}$. While in Ref. \onlinecite{DiffusiveRenyi} we argued that this bound is not tight, and $S_2^{(a)}$ behaves qualitatively differently from $S_1$ for \emph{homogenous} initial states, it is nevertheless interesting to consider the behavior in the former for \emph{inhomogenous} states. As we now show, several of the features predicted by our surface growth model in Sec.~\ref{sec:nofluc} appear also in the dynamics of $S_2^{(a)}$.

We once again start by considering a domain wall initial state. As argued before in Sec.~\ref{sec:inhom}, since the charge density obeys the diffusion equation, it has a profile $\braket{\hat Q_x(t)}$  that depends only on the combination $x/\sqrt{Dt}$, where for the present case the diffusion constant is $D=1/2$. Assuming local equilibration, i.e., that the on-site reduced density matrix takes the form $\rho_x(t) \propto e^{-\mu(x,t)\hat Q_x}$ for some local chemical potential $\mu(x,t)$ (which, as we show in App.~\ref{app:local_eq}, indeed holds after some short-time relaxation process), one can use subadditivity to upper bound for the von Neumann entropy as
\begin{equation}\label{eq:dw_subadd}
S_1(x,t) \leq \sum_{x'\leq x} s(x',t) = \sqrt{Dt}\,f\left(\frac{x}{\sqrt{Dt}}\right),
\end{equation}
where $s(x',t)$ is the one-site von Neumann entropy of site $x'$ at time $t$. Therefore the von Neumann entropy cannot grow faster than $\sqrt{t}$. Clearly the same upper bound applies to all higher index R\'enyi entropies, as well as to $S_2^{(a)}$.

\begin{figure}
	\includegraphics[width=1.\columnwidth]{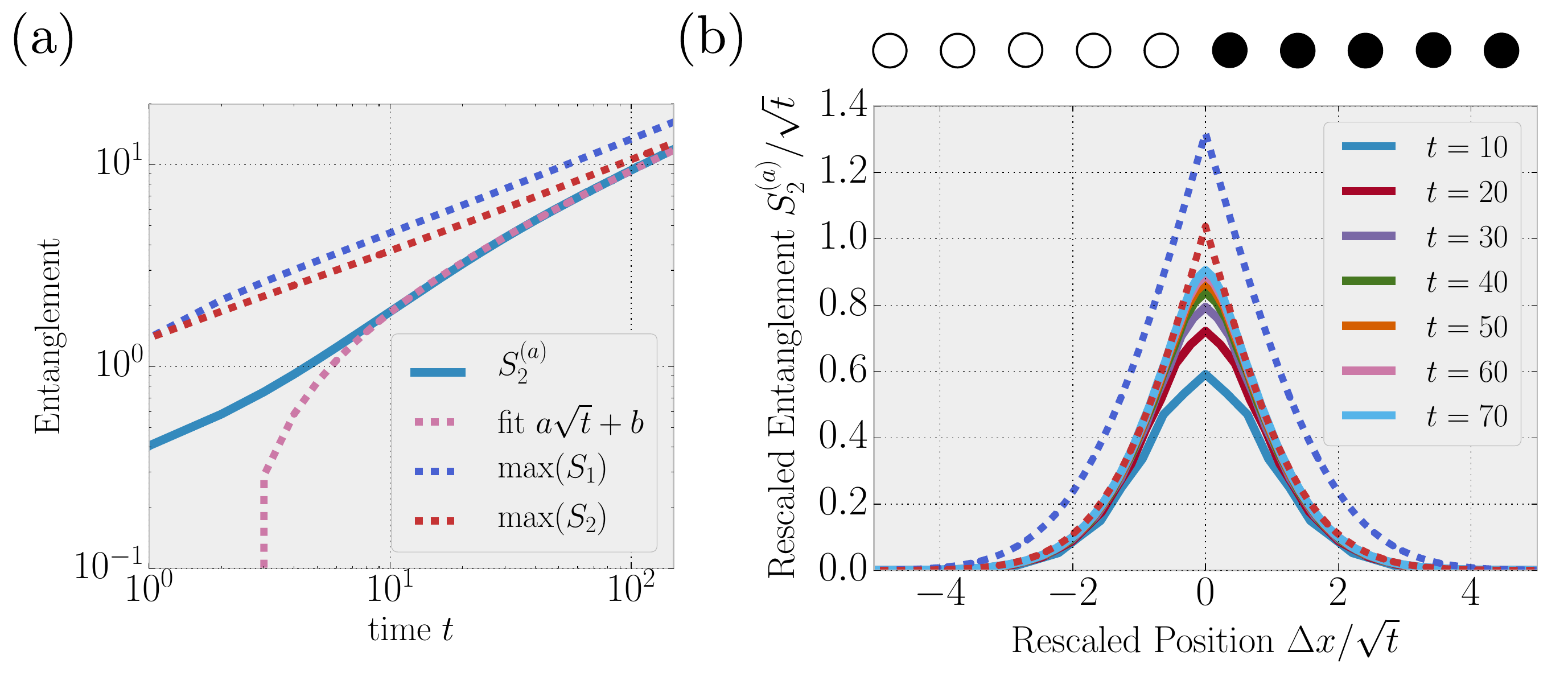}
\caption{Entanglement growth for an initial domain wall in the random circuit with $N=1$. The solid lines are data obtained from numerically evaluating the annealed average of $S_2$ which provides a lower bound for $\overline{S_{1}}$ and $\overline{S_{2}}$. (a) Growth of entanglement in the middle of the domain wall is well described by $\sim \sqrt{t}$ asymptotically. For the numerical computation we used bond dimension $\chi = 5000$ and a chain of $L=400$ sites. The dashed blue line is the upper bound from Eq.~\eqref{eq:dw_subadd}, given by the cumulative sum of the on-site von Neumann entropies associated to the average local charge density. The red dashed line is the cumulative sum of the local second R\'enyi entropy, which gives a good approximation of the behavior of the annealed average at long times. (b) shows the profile of entanglement over the chain ($L=200$, $\chi=4000$), which at long times exhibits a scaling collapse when both the position $x$ and the entanglement $S_2^{(a)}$ are rescaled by $\sqrt{t}$. Dashed lines again indicate the cumulative sum of on-site entropies.} 
	\label{fig:dw_N=1}
\end{figure}

The annealed average, $S_2^{(a)}$, on the other hand, provides a \emph{lower bound} for both $\overline{S_1}$ and $\overline{S_2}$. Therefore if $S_2^{(a)}$ also grows as $\propto \sqrt{t}$ then it follows that all of these quantities have to have the same diffusive growth. This is exactly what we find by evaluating the annealed average numerically: after some initial short-time dynamics its behavior at $x=0$ is well described by $S_2^{(a)}(0,t) = a\sqrt{t} + b$ for some constants $a,b$ as shown in Fig.~\ref{fig:dw_N=1}. In fact we find that the profile of $S_2^{(a)}(x,t)$ is well approximated by the sum of the second R\'enyi entropies of the local one-site density matrices, $\sum_{x' \leq x} s_2(x,t)$, where $s_2(x,t) = -\log(1 -2\mathfrak{q} + 2\mathfrak{q}^2)$ is the R\'enyi entropy density associated to a state in local equilibrium, parametrized by the charge density $\mathfrak{q} = \mathfrak{q}(x,t) \equiv \overline{\braket{\hat Q_x(t)}}$. That is, despite the fact that $S_2$ does not obey subadditivity, its annealed average behaves as if it satisfied a modified version of the inequality in Eq.~\eqref{eq:dw_subadd}, where we replace the von Neumann entropy density with that of the second R\'enyi entropy. This interpretation is made plausible by a random circuit calculation in App.~\ref{app:partfunc}, which shows that when charge fluctuations are ignored, $S_2^{(a)}$ obeys an equation similar to Eq.~\eqref{eq:large_N_eom}, but with a modified constant term.

\subsection{Deterministic Floquet spin chain}\label{sec:floquet_def}

While random circuits have many advantages, we also wish to compare our predictions to deterministic models of unitary dynamics. While the KPZ fluctuations mentioned in Sec.~\ref{sec:nofluc} are particular to noisy dynamics, we expect that the imprints of charge transport, such as the $\sqrt{t}$ growth of entanglement for a domain wall, do generalize to the deterministic setting. We expect this to be the case since many of the arguments discussed above largely follows from the diffusive dynamics of conserved quantities, which is expected to hold generically at high temperatures even in deterministic systems~\cite{BLOEMBERGEN1949,DEGENNES1958,KADANOFF1963,Rosch13,Bohrdt16}.

To this end we also perform calculations in a Floquet (periodically driven) spin chain, introduced by the authors in Ref. \onlinecite{OTOCDiff1}, where it was shown to exhibit clear diffusive tails. Considering such a driven system has the advantage of simplicity; U(1) spin is the only conserved quantity. We leave it to future studies to consider other models, in particular those with energy conservation, and the appropriate generalizations of CDWs and domain walls in that context. 

The model consists of a spin-$\frac{1}{2}$ chain evolved by a time-periodic Hamiltonian. A single driving sequence consists of four parts, with the so-called Floquet unitary given by
\begin{align}\label{eq:Floquet_def}
U_\text{F} &= e^{-i\tau H_4} e^{-i\tau H_3} e^{-i\tau H_2} e^{-i\tau H_1} \nonumber \\
H_1 &= J_{z}^{(1)} \sum_r \hat Z_r \hat Z_{r+1} \nonumber \\
H_3 &= J_{z}^{(2)} \sum_r \hat Z_r \hat Z_{r+2} \nonumber \\
H_2 &= H_4 =  J_{xy} \sum_r \left(\hat X_r \hat X_{r+1} + \hat Y_r \hat Y_{r+1}\right),
\end{align}
where $\hat X_r, \hat Y_r, \hat Z_r$ are Pauli spin operators on site $r$. Every part of the drive individually conserves the spin z component, $[H_a,\sum_r \hat Z_r] = 0$ for $a=1,2,3,4$. We take the period time to be $T \equiv 4\tau = 1$ and the couplings to be all order 1, namely $J_z^{(1)} = (\sqrt{3} + 5) / 6$, $J_z^{(2)} = \sqrt{5} / 2$ and $J_{xy} = (2 \sqrt{3} + 3) / 7$. 

\begin{figure}
	\includegraphics[width=0.7\columnwidth]{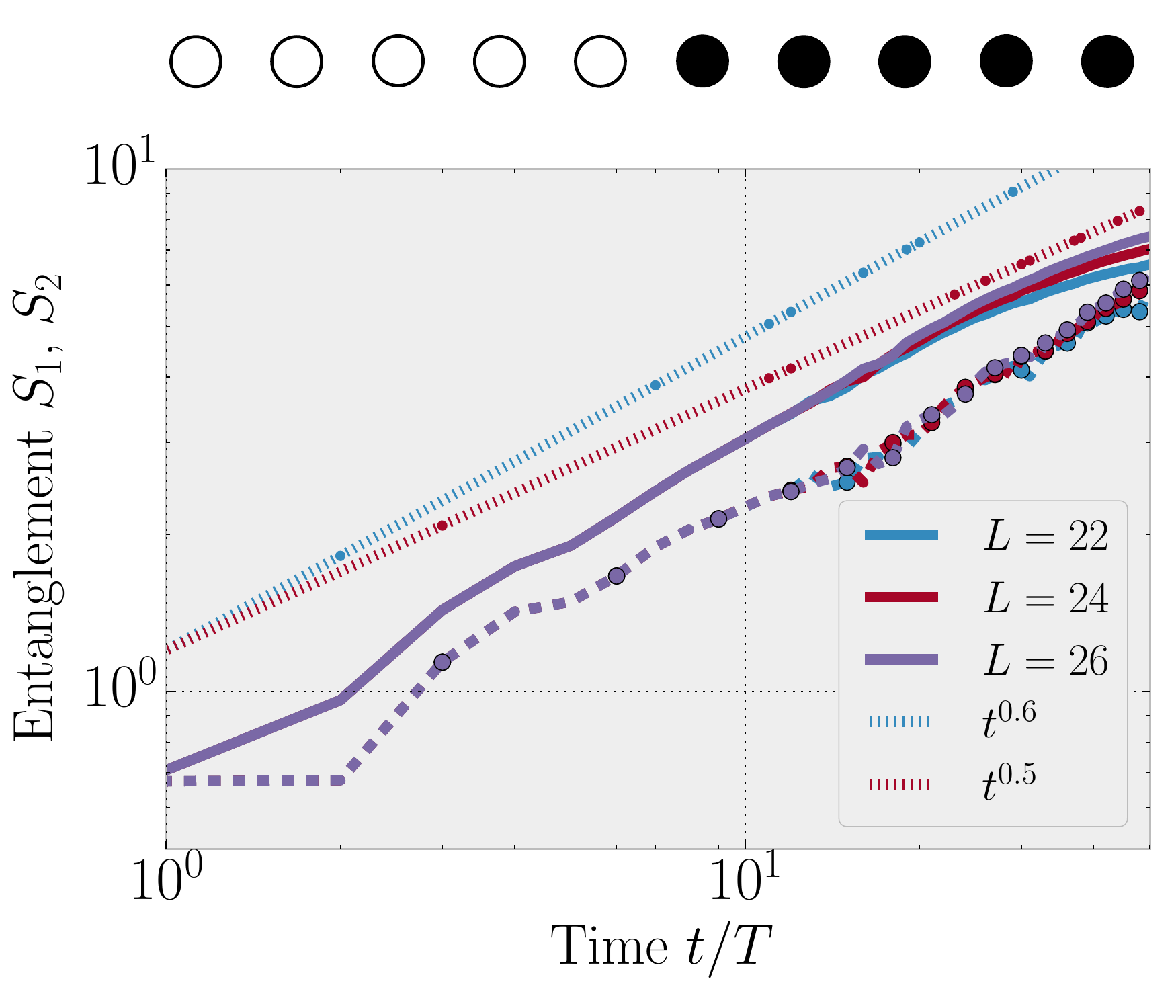}
	\caption{Entanglement growth between two halves of a domain wall, evolved with the Floquet unitary~\eqref{eq:Floquet_def}. Solid lines represent the von Neumann entropy $S_1$ while dashed lines are the second R\'enyi $S_2$. Computed using exact diagonalization for system sizes $L=22,24,26$.} 
	\label{fig:Floquet_dw_ed}
\end{figure}

As emphasized in the previous section, the scaling observed for domain wall-like initial states, where both the size and the width of the features in the entanglement profile grow as $\propto\sqrt{t}$ is a direct consequence of diffusive transport - the upper bound derived above only relies on the assumptions of a) local equilibration and b) diffusive charge-transport, both of which are expected to apply to generic many-body systems at high temperatures. While this is only an upper bound, and in principle slower than $\sqrt{t}$ growth is possible, we do not expect this to be the case unless there are some further constraints on the dynamics~\footnote{Such is the case of integrable systems, which can exhibit logarithmic growth of the entanglement~\cite{Alba2014,Vidmar2017}, even though their transport properties are typically \emph{faster} than diffusive.}. Indeed, evolving a domain wall initial state with the Floquet unitary~\eqref{eq:Floquet_def} we find that the entanglement entropy across the middle of the chain grows slower than ballistically, approximately as $t^{0.5-0.6}$~\footnote{Note that even in the random circuit results at $N=1$, shown in Fig.~\ref{fig:dw_N=1}, the $~\sqrt{t}$ growth is only true asymptotically and at short times a slightly faster growth is observed.}. This result, shown in Fig.~\ref{fig:Floquet_dw_ed}, reinforces our expectation that the asymptotic growth should be $\propto \sqrt{t}$.

Considering the full entanglement profile, however, we do not observe the scaling collapse seen previously for the random circuit (i.e. a $\propto \sqrt{t}$ width). Our interpretation for this is the following. While the deterministic model we consider is expected to show diffusive transport at `high temperatures', or in this case at filling fractions close to $1/2$, this is not the case when the filling is very small (or very close to $1$), which is the case far from the middle of the domain wall. In these regions there are very few particles (or very few holes), which therefore propagate without many scattering events. This suggest that in the deterministic model there should be two distinct regimes: near the origin where the domain wall has sufficiently melted and the average density is close to $1/2$ the diffusive scaling should apply, while at the tails of the domain wall, where the filling is close to (but different from) $0$ or $1$ there is a ballistic region. We show some further data in support of this interpretation in App~\ref{app:Floquet_dw_app}.

\section{Logarithmic growth of number entanglement}\label{sec:charge_entropy}

Before concluding, let us briefly comment on the dynamics of the von Neumann entropy in charge-conserving systems in the absence of large-scale inhomogeneities. The discussion that follows applies to initial states that have a fixed amount of total charge in the entire system. A simple example of such a state is a N\'eel state, where exactly every second site is occupied. While this state is not completely homogenous, its charge-density becomes uniformly $1/2$ after some quick local equilibration process. The resulting state, however, is still far from \emph{global} equilibrium, as indicated, among other things, by the fluctuations in the conserved charge. Consider a subsystem $A$, containing $L_A$ sites, and the total charge contained in it, as measured by the operator $\hat Q_A \equiv \sum_{x\in A} \hat Q_x$. In an equilibrium state, $\rho_A \propto e^{-\mu \hat Q_A}$ the charge variance should obey a volume law, $\text{Var}(Q_A) \equiv \braket{\hat Q_A^2} - \braket{\hat Q_A}^2 \propto L_A$. As suggested by our arguments made previously in Ref. \onlinecite{DiffusiveRenyi} for the random circuit model, as well as those established more generally in the literature~\cite{Rosch13}, to leading order the evolution of the variance should be governed by diffusion, indicating that it takes a time $\propto L_A^2$ for it to reach its final value. Our goal is to understand how this diffusively slow relaxation affect the dynamics of entanglement.

To see the effects of charge diffusion more clearly it is worthwhile to write the von Neumann entropy as a sum of two contributions, one of which is related directly to the distribution of total charge in the subsystem. In order to achieve this, note that the conservation of a fixed total charge in the system implies that the reduced density matrix has a block-diagonal structure 
\begin{equation}
\rho_A = \sum_{Q_A} p_{Q_A} \rho_A^{(Q_A)},
\end{equation}
where $\rho_A^{(Q_A)}$ corresponds to the block with total charge $Q_A$ in subsystem $A$, normalized to have unit trace. With this normalization, the prefactor $p_{Q_A}$ is exactly the probability of having a total number of $Q_A$ charges in $A$, given by the expectation value $\text{tr}(\rho_A \hat P_{Q_A})$ where $\hat P_{Q_A}$ is the projector appearing in the spectral decomposition $\hat Q_A = \sum_{Q_A} Q_A \hat P_{Q_A}$.

With this block-decomposition in hand one can rewrite the von Neumann entropy as~\cite{Lukin18}
\begin{align}\label{eq:split_vN}
S_1[\rho_A] &= -\sum_{Q_A} p_{Q_a} \log{p_{Q_A}} - \sum_{Q_A} p_{Q_A} \text{tr}\left(\rho_A^{(Q_A)} \log{\rho_A^{(Q_A)}} \right) \nonumber \\
&= S_Q + S_\text{conf}.
\end{align}
The first term is the entropy associated to the probability distribution $p_{Q_A}$, which we will refer to as the \emph{number entropy}. The second term is the average entropy of the blocks which, following Ref. \onlinecite{Lukin18}, we call the configurational entropy, as it is associated to superpositions between different ways of arranging the same number of total charges within the subsystem. Importantly, the number entropy is independently measurable in cold atom experiments~\cite{Lukin18}.

\begin{figure}
	\includegraphics[width=1.\columnwidth]{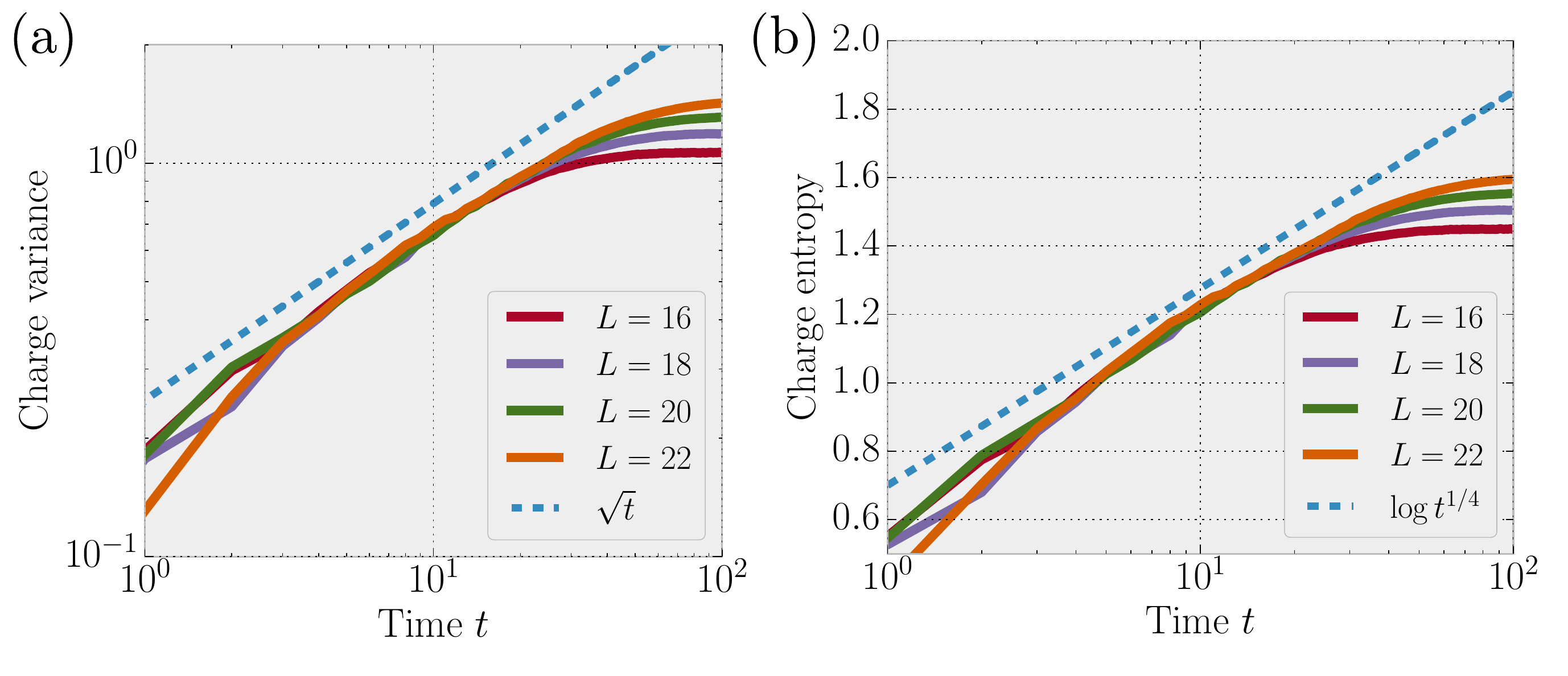}
	\caption{Evolution of the probability distribution of having total charge $Q$ in the left half of a spin chain, initialized in a N\'eel state and evolved with the random unitary circuit. a) The charge variance $\text{Var}(Q)$ grows as $\propto\sqrt{t}$ and saturates to a volume law. b) The number entropy $S_Q$ grows as the logarithm of the standard deviation, $S_Q = \log{t^{1/4}} + const.$ and saturates to a value $\log{L^{1/2}} + const.$} 
	\label{fig:charge_fluc_Neel}
\end{figure}

For the types of initial states considered here, with roughly homogeneous charge-distributions, the mean value of the distribution $p_{Q_A}$ is close to its equilibrium value even after short times. However, this distribution is initially very narrow, with a variance $\text{Var}(Q_A)$ that should grow as $\propto\sqrt{t}$, due to diffusion, as argued in Ref. \onlinecite{DiffusiveRenyi}. This provides an upper bound on the number entropy, as $S_Q \leq \ln\left(\sqrt{2\pi e \text{Var}(Q_A)} \right) \approx \log{t^{1/4}} + \text{constant}$. We indeed confirm that the growth of $S_Q$ has this form, by performing the exact random circuit time evolution for a small system, as shown in Fig.~\ref{fig:charge_fluc_Neel}. Note that approximating the entropy as the logarithm of the standard deviation is also consistent with the long-time saturation value, which should be $S_Q(t\to\infty) = \log{\sqrt{L_A}} +  \text{const.}$ for the binomial distribution. As emphasized earlier, $S_Q$ itself is a measurable quantity and therefore this prediction could in principle be confirmed in experiments. We also expect this to manifest as a logarithmically growing component of the full von Neumann entropy~\eqref{eq:split_vN}, although in principle it is possible that this contribution is cancelled by some part of the configurational entropy.

\section{Discussion}\label{sec:discuss}

We investigated the dynamics of entanglement in situations where a one-dimensional system approaches equilibrium from an initial state with large scale inhomogeneities in some conserved quantity, such as particle number. We argued that all other things being equal, entanglement grows more slowly at very low and very high fillings: in particular, entanglement does not grow in empty regions, leading to a bottleneck which strongly limits the overall global growth of entropy. We made this intuition more precise by noting (using subadditivity) that the entanglement growth at each time step can be bounded by the local entropy density, which in turn is related to the local filling/chemical potential, once the system has locally equilibrated. Using this subadditivity bound as a heuristic equation of motion for the von Neumann entropy under a random circuit evolution \eqnref{eq:kpz_diff}, we constructed a stochastic surface growth model for the von Neumann entropy, resembling a noisy Kardar-Parisi-Zhang equation coupled to a diffusively evolving charge distribution. This can be recast as a modified minimal cut variational principle similar to that of Ref. \onlinecite{JonayNahum}, which suggests generalizations both to deterministic systems, and to higher dimensions. 

We investigated the behavior of the surface growth equation of motion for various initial states. We showed that domain wall-like jumps in the initial filling lead to diffusively growing features in the entanglement profile. When such domain walls are combined into a N\'eel-like charge density wave of wavelength $\lambda$, this leads to an inhomogenous entanglement profile, with peaks at the positions of the initial domain walls. These peaks grow (and spread in space) as $\sim \sqrt{t}$ until the CDW melts at times $\sim \lambda^2$. The entropy profile eventually loses memory of the initial charge distribution, but only on a parametrically longer time scale $\sim \lambda^2 (1+ \log(\lambda))$. 

Our surface growth model predictions were supported by a circuit averaged calculation of the annealed second R\'enyi entropy (which bounds below the von Neumann entropy). Furthermore we have demonstrated that the deterministic, non-noisy model of \eqnref{eq:Floquet_def} also exhibits the diffusive growth of entanglement for the case of a single domain wall, (albeit with some additional subtleties related to regions with very low density of excitations where transport becomes ballistic). Finally, motivated by recent experiments \cite{Lukin18}, we investigated the dynamics of the so-called number entanglement, and argued that it grows logarithmically in time in systems with diffusive transport. Our argument relied on the assumption that the charge variance of extensive regions tends to grow as $\propto \sqrt{t}$, an assumption based on general hydrodynamic considerations and verified by a random circuit calculation.

Our work suggests further avenues for research. It would be interesting to see whether our heuristically motivated equation of motion for the entanglement can be put on firmer ground in the random circuit setting, perhaps using the methods of Ref. \onlinecite{ZhouNahum}. Furthermore, while a generalization of our model to higher dimensions, through the minimal cut/membrane picture, appears natural, establishing it more rigorously, and investigating its properties remains an open problem. Another interesting avenue is understanding entanglement growth in situations where there is an interplay between particle/spin transport and energy conservation. For example, the dynamics of charge domain walls in U(1)-symmetric \emph{Hamiltonian} systems, can be more complex, due to such interplay constraining the domain wall motion. Investigating entanglement dynamics in this case, for generic non-integrable systems is an interesting open question. 

Lastly, it would be interesting to pin down the connection between the minimal cut formalism and holographic calculations \cite{Mezei18}; do the diffusive characteristics of entropy growth show up in analogous holographic quench calculations? 

\acknowledgements{We thank Herbert Spohn, Adam Nahum and Juah Garrahan for useful discussions. FP acknowledges the support of the DFG Research Unit FOR 1807 through grants no. PO 1370/2- 1, TRR80, the Nanosystems Initiative Munich (NIM) by the German Excellence Initiative, the Deutsche Forschungsgemeinschaft (DFG, German
Research Foundation) under Germany's Excellence Strategy -- EXC-2111-390814868 and the European Research Council (ERC) under the European Unions Horizon 2020 research and innovation program (grant agreement no. 771537). CvK is supported by a Birmingham Fellowship.}

\appendix

\section{Mapping average purity to classical partition function}\label{app:partfunc}

As noted in the main text, the simplest entanglement measure which we can calculate tor our random circuit models is the annealed average of the second R\'enyi entropy, which is given by the logarithm of the average purity $\mathcal{P}$,
\begin{equation}
S_2^{(a)}(t) \equiv -\log{\overline{\tr(\rho_A(t)^2)}} = -\log{\overline{\mathcal{P}[\rho_A(t)]}}.
\end{equation}
Since this is a quantity that only involves four instances of the time evolution operator $U$, the averaging over a single gate of the circuit can be done in a relatively straightforward manner, resulting in a 2D tensor network representation of the average purity. The boundary conditions of this partition function at times $0$ and $t$ depend on the initial state $\rho(0)$ and on the choice of the subsystem $A$, respectively. This representation allows us to efficiently evaluate $S_2^{(a)}(t) $, by contracting the 2D tensor network via the time-evolving block decimation algorithm~\cite{VidalTEBD}, resulting in the data shown in Sec.~\ref{sec:dw_N=1} and below in App.~\ref{app:local_eq}. The same mapping has been developed in Ref. \onlinecite{OTOCDiff2} (generalizing the calculation of Ref. \onlinecite{Nahum16} to the U(1)-symmetric case) to calculate other quantities, and we summarize it briefly below. 

The basic building block of the 2D partition function is a 4-leg tensor, which we obtain by averaging over four copies of a single unitary gate $U$. The operator $U$ acts on a Hilbert space $\mathcal{H}^2$ (i.e. the Hilbert space of two neighboring sites of the chain), which can be decomposed into sectors with fixed total charges as $\mathcal{H}^2 = \bigoplus_Q \mathcal{H}^2_Q$, each of which has a dimension $d_Q$. The effective tensor associated to this gate then has the following form~\cite{OTOCDiff2}:
\begin{multline}\label{eq:onegate_Curtbasis}
\overline{U^* \otimes U \otimes U^* \otimes U} = \sum_{Q_1,Q_2} \frac{1}{d_{Q_1}d_{Q_2}} \ket{\mathcal{I}_{Q_1Q_2}}\bra{\mathcal{I}_{Q_1Q_2}} + \\
\sum_{Q_1,Q_2} \frac{1}{d_{Q_1}d_{Q_2} - \delta_{Q_1Q_2}} \ket{\mathcal{J}_{Q_1Q_2}}\bra{\mathcal{J}_{Q_1Q_2}},
\end{multline}
where the states $\ket{\mathcal{I},\mathcal{J}}$ on the four-copy Hilbert space are defined as
\begin{align}
\ket{\mathcal{I}_{Q_1Q_2}} &\equiv \sum_{\alpha \in \mathcal{H}^2_{Q_1}} \sum_{\beta \in \mathcal{H}^2_{Q_2}} \ket{\alpha\alpha\beta\beta}\nonumber \\
\ket{\mathcal{J}_{Q_1Q_2}} &\equiv \sum_{\alpha \in \mathcal{H}^2_{Q_1}} \sum_{\beta \in \mathcal{H}^2_{Q_2}} \left[ \ket{\alpha\beta\beta\alpha}- \frac{\delta_{Q_1Q_2}}{d_{Q_1}}  \ket{\alpha\alpha\beta\beta} \right].
\end{align}
These states belong to four copies of the total (two-site) Hilbert space. In order to contract the tensor network one then needs to split them up to states living on four copies of a single-site Hilbert space, which in principle can be done in a many different ways, depending on the choice of local basis. 

Each unitary gate therefore maps to such a 4-leg tensor, which then need to be contracted, as defined by the geometry of the circuit (e.g. the `brick wall' geometry of Fig.~\ref{fig:circuit_def}(c)). The final step in calculating the average purity is to specify the boundary conditions. We note that the purity of subsystem $A$ can be rewritten as~\cite{HastingsSwap,DemlerEntanglement,Hayden2016,Nahum17,DiffusiveRenyi}
\begin{equation}\label{eq:purity_twocopies}
\mathcal{P}[\rho_A] = \text{tr}_A(\rho_A^2) \equiv \text{tr} (\mathcal{S}_A [\rho \otimes \rho]),
\end{equation}
where $\rho \otimes \rho$ is two copies of the density matrix of the \emph{whole} system, and $\mathcal{S}_A$ is an operator on this doubled Hilbert space that swaps the two copies inside subsystem $A$ and acts as the identity on its complement. The trace on the right hand side should be understood as a sum over a complete set of states in the two-copy Hilbert space. The boundary conditions of the 2D tensor network representation therefore correspond to $\rho(0) \otimes \rho(0)$ on the boundary at time $0$ and by $\mathcal{S}_A$ on the boundary at time $t$. In particular, denoting by $\mathcal{H}^1$ the single site Hilbert space, the latter reads
\begin{equation}
    \mathcal{S}_A = \bigotimes_{x\in A} \ket{\mathcal{I}}_x \, \bigotimes_{x'\notin A} \ket{\mathcal{K}}_{x'},
\end{equation}
where
\begin{align}
    \ket{\mathcal{I}} \equiv \sum_{\alpha, \beta \in \mathcal{H}^1} \ket{\alpha\alpha\beta\beta} & & \ket{\mathcal{K}} \equiv \sum_{\alpha, \beta \in \mathcal{H}^1} \ket{\alpha\beta\beta\alpha}.
\end{align}
The other boundary condition depends on the initial state $\rho(0) = \ket{\psi(0)} \bra{\psi(0)}$. For product states in the local charge basis, relevant in the present context, the two on-site states that appear are $\ket{0000}$ and $\ket{1111}$.

\subsection{Equation of motion for well defined local charge}

Apart from providing an efficient way of numerically evaluating $S_2^{(a})(t)$ in the random circuit, utilized in Sec.~\ref{sec:dw_N=1}, the above calculation can also be used to derive an update rule in the case when a 2-site unitary is acting on a bond with fixed amount of charge, similarly to what has been done for the Hartley entropy in Sec.~\ref{sec:Hartley}. Here we show that this leads to an update rule similar to Eq.~\eqref{eq:large_N_eom}, but with a modified constant term, reflecting the differences between the von Neumann and second R\'enyi entropies.

For the purposes of understanding the effect of a single gate, it is enough to look at the two-site version of the swap operator, i.e. $\mathcal{I} \otimes \mathcal{K}$. Applying the operator~\eqref{eq:onegate_Curtbasis} to this state we get two terms, which read
\begin{align}\label{eq:large_N_twoterms}
\sum_{Q_1,Q_2} \frac{\braket{\mathcal{I}_{Q_1Q_2}|\mathcal{I}\mathcal{K}}}{d_{Q_1}d_{Q_2}} \ket{\mathcal{I}_{Q_1Q_2}} &= \sum_{Q_1,Q_2} \frac{\eta f_{Q_1Q_2} \ket{\mathcal{I}_{Q_1Q_2}}}{d_{Q_1}d_{Q_2}}; \nonumber \\
\sum_{Q_1,Q_2} \frac{\braket{\mathcal{J}_{Q_1Q_2}|\mathcal{S}\mathcal{I}}}{d_{Q_1}d_{Q_2}} \ket{\mathcal{I}_{Q_1Q_2}} &= \sum_{Q_1,Q_2} \frac{\eta f_{Q_1Q_2} \ket{\mathcal{J}_{Q_1Q_2}}}{d_{Q_1}(d_{Q_2}+\delta_{Q_1Q_2})},
\end{align}
where we have defined
\begin{equation}
f_{Q_1Q_2} \equiv \sum_{a,b_1,b_2=0}^{N} \delta_{a+b_1=Q_1}\delta_{a+b_2=Q_2} \frac{\eta_a\eta_{b_1}\eta_{b_2}}{\eta},
\end{equation}
in terms of the dimension of the charge $a$ sector, $\eta_a \equiv \dim \mathcal{H}^1_Q$, and the total single site dimension $\eta \equiv \sum_a \eta_a$.

To arrive at the equation of motion we are going to assume that the local charge on the two sites in question is exactly $\bar{Q}$, so that
\begin{equation*}
P_Q \rho = \delta_{Q\bar Q} \rho,
\end{equation*}
where $P_Q$ is the operator that projects onto the charge $Q$ sector of the two-site Hilbert space. Therefore, when evaluated on a density matrix that satisfied the above relation, the sum of the two terms in Eq.~\eqref{eq:large_N_twoterms} can be replaced with
\begin{equation}
\frac{\eta  f_{\bar Q \bar Q}}{d_{\bar Q}(d_{\bar Q}+1)} \left( \ket{\mathcal{I}\mathcal{I}} + \ket{\mathcal{K}\mathcal{K}} \right).
\end{equation}
The right hand side is therefore proportional to the sum of the `purity superoperators' at the two neighboring bonds. The last remaining step is to evaluate the prefactor. In the large-$N$ limit it simplifies to $\frac{\eta  f_{\bar Q \bar Q}}{d^2_{\bar Q}}$. A short calculation reveals that this quantity is just the one-site purity associated with the maximally entangled two site density matrix with charge exactly $\bar Q$, namely $\rho = P_{\bar Q} /d_{\bar Q}$.

Using the assumption of well defined local charge, along with the large-$N$ limit, we therefore arrive at a closed equation of motion for the Purity $\mathcal{P}(x,t)$ under the effect of a local 2-site gate at position $x$, which reads
\begin{equation}\label{eq:purity_large_M}
\mathcal{P}(x,t+1) = e^{s_2(\bar Q)} \left(\mathcal{P}(x-1,t) + \mathcal{P}(x+1,t)\right).
\end{equation}
Here $s_2(\bar Q) \equiv \ln\left(\frac{\eta  f_{\bar Q \bar Q}}{d^2_{\bar Q} }\right)$ is the density of second R\'enyi entropy, associated to the local charge density on the bond $x,x+1$. Taking the logarithm of Eq.~\eqref{eq:purity_large_M} (with base $2^N$) and once again making use of the large-$N$ limit results in a very simple update rule for the annealed average second R\'enyi entropy under the effect of a 2-site gate:
\begin{equation}\label{eq:large_M_eom}
S_2^{(a)}(x,t+1) = \text{min}(S_2^{(a)}(x-1,t),S_2^{(a)}(x+1,t)) + s_2(\bar Q).
\end{equation}
This has the same form as the update rule~\eqref{eq:large_N_eom} we used in our effective surface growth model for the von Neumann entropy, except for the fact that the local density of von Neumann entropy has been replaced by the local density of $S_2$. This update rule is consistent with our numerical results in Sec.~\ref{sec:dw_N=1}, where we observed that the profile of $S_2^{(a)}$ starting from a domain wall at long times behaves as if it obeyed sub-additivity. This is despite the fact that the update rule~\eqref{eq:large_M_eom} clearly misses the important effect of charge fluctuations for homogenous states, uncovered in Ref. \onlinecite{DiffusiveRenyi}.

\section{Further numerical results on spin-$\frac{1}{2}$ chains}\label{app:numerics}

In this appendix we gather some further data on inhomogenous quenches, both in the random circuit and in the Floquet model, to complement the results shown in the main text.

\subsection{Local equilibration in the spin-$\frac{1}{2}$ random circuit}\label{app:local_eq}

\begin{figure}[t!!]
	\includegraphics[width=1.\columnwidth]{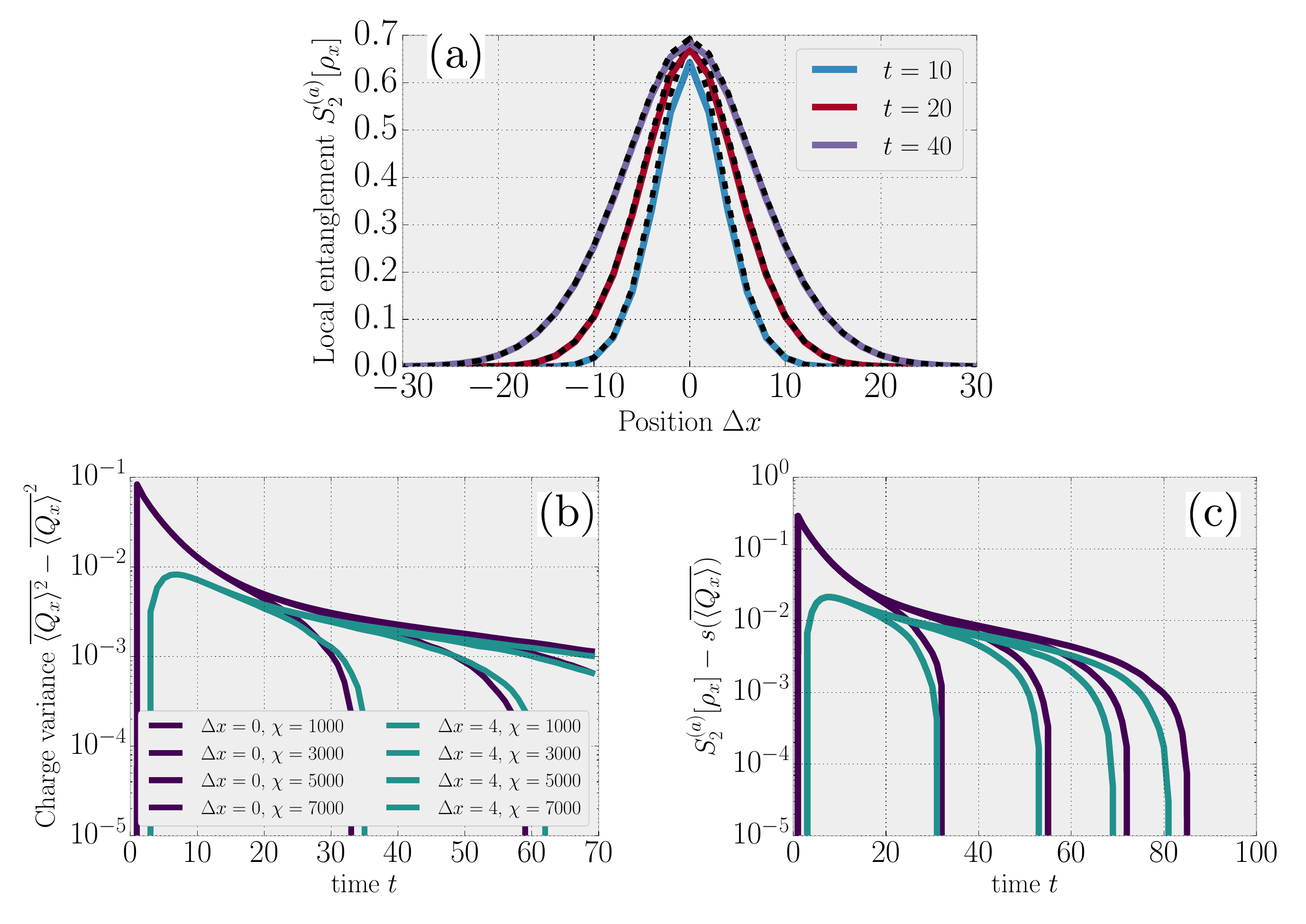}
	\caption{Local equilibration for a domain wall state in the random circuit with $N=1$. (a) The local purity entropy on average is well approximated by the purity of the local equilibrium state associated to the average charge. (b) The distance to the equilibrium value for two different positions as a function of time for different bond dimensions $\chi$ shows a decay to zero. (c) the size of the fluctuations between the local charge densities in different circuit realizations (as captured by their variance) similarly decrease in time. } 
	\label{fig:dw_local_eq}
\end{figure}

Our general argument, putting an upper bound proportional to $\sqrt{t}$ on the growth of von Neumann entropy for a domain wall, relied on the notion of local equilibration. Here we check numerically that this assumption indeed holds in the spin-$\frac{1}{2}$ local random circuit we considered in the main text. This consists of two parts. First of all, in order for local equilibration to $\overline{\braket{\hat Q_x(t)}}$ to make sense, the circuit-to-circuit fluctuations in $\braket{\hat Q_x(t)}$ around this average should become small. This means that we can assign a local charge to each site, which should be approximately the same for all realizations of the circuit, making the right hand side of Eq.~\eqref{eq:dw_subadd} well-defined. Second, we need the one-site density matrices to be close to the Gibbs state defined by this local charge density. As we show in Fig.~\ref{fig:dw_local_eq} both of these requirements are satisfied at sufficiently long times.

To check the first requirement, we compute numerically the statistical variance of the local charge in the random circuit, i.e. the variance between circuit realizations of the quantum expectation value $\braket{\hat Q_x}$. This can be computed as a 2D partition function, similarly to the calculation of $S_2^{(a)}$ outlined in App.~\ref{app:partfunc}. We find that the variance decays in time, as shown by Fig.~\ref{fig:dw_local_eq}(b), indicating that at long times it is meaningful to consider locally equilibrated states depending on a circuit-independent average local charge. 

To check if the on-site density matrix is indeed locally equilibrated, we compare its annealed average second R\'enyi entropy with the entropy density expected for an eqilibrium state with the same charge density. Note that since the on-site density matrix only has two parameters, one of which is fixed by the charge density, the equivalence of R\'enyi entropies is sufficient to establish local equilibrium. We indeed find that the local R\'enyi entropy tends to its equilibrium value at long times; this is shown in Fig.~\ref{fig:dw_local_eq}(c). In both cases we show results for different bond dimensions $\chi$ used in the evaluation of these quantities as 2D tensor networks to indicate where the results are converged.

\subsection{Entanglement profile in the Floquet model}\label{app:Floquet_dw_app}

\begin{figure}[h!]
	\includegraphics[width=1.\columnwidth]{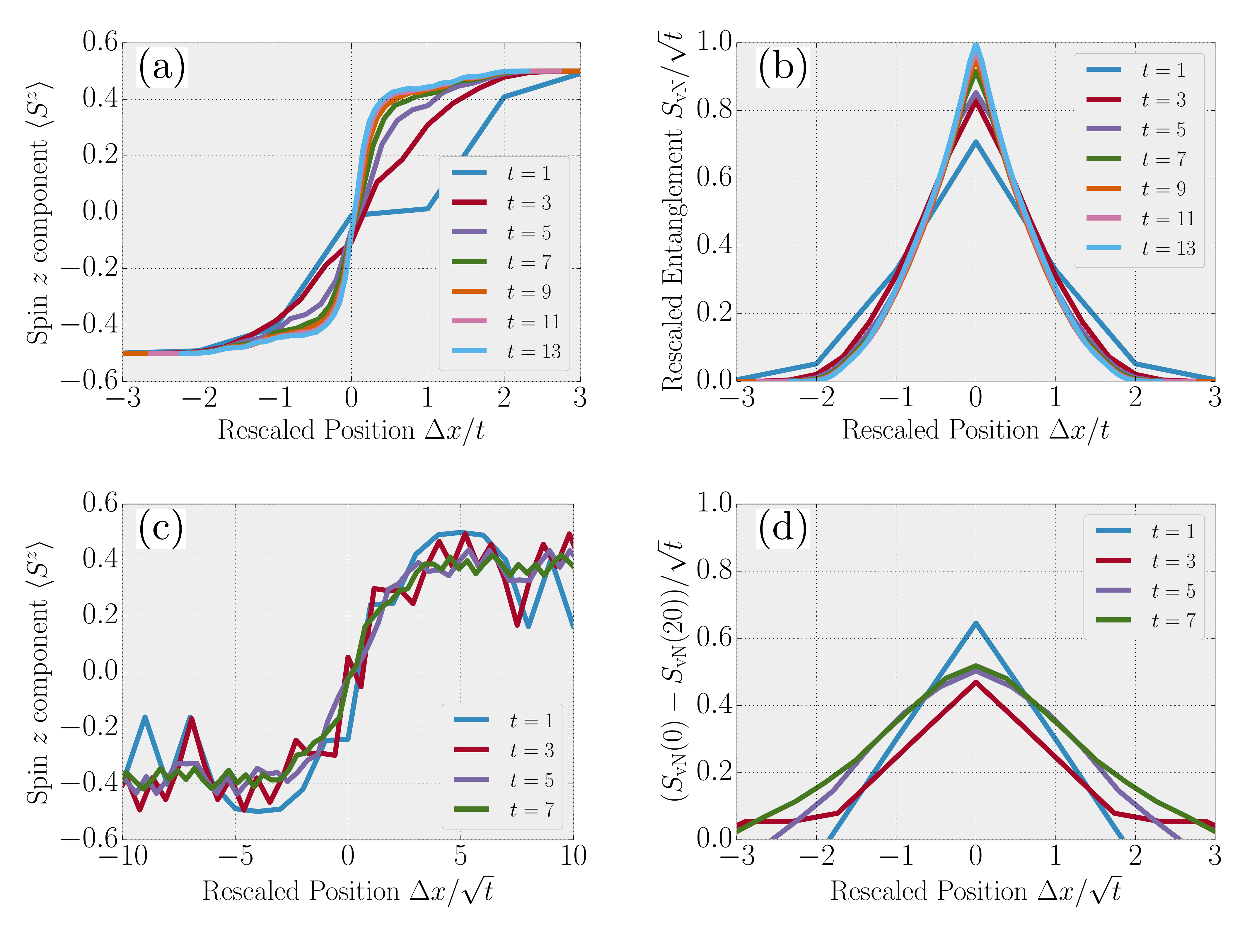}
	\caption{Spin (left) and entanglement profiles (right) at different times for the Floquet model~\eqref{eq:Floquet_def}, for a domain wall (upper row) or a generalized domain wall (lower row). In the generalized domain wall the average densities are $n_L = 1/4$ and $n_R = 3/4$ on the two sides respectively. The domain wall has ballistically spreading tails in the low density regions while the generalized domain wall is diffusive everywhere.} 
	\label{fig:Floquet_dw_profiles_tebd}
\end{figure}

As stated in Sec.~\ref{sec:floquet_def} of the main text, while we do observe the expected sub-ballistic growth for the half-chain entanglement of a domain wall initial state in the Floquet spin chain model, we find that the spatial width of the entanglement profile spreads faster than $\sqrt{t}$. We interpreted this by noting that at in the tails of the melting domain wall, far on the left (right) from its center, the conserved spin density is close to minimal (maximal). In these regions, a weakly-interacting quasi-particle picture should therefore be approptiate, implying ballistic spreading. This is supported by considering the spatial profile of the charge, which shows these two distinct regimes: tails that spread linearly and a middle region which spreads out more slowly, as shown in the upper panels of Fig.~\ref{fig:Floquet_dw_profiles_tebd}

Based on the above interpretation it is expected that the diffusive (spatial) spreading of the entanglement profile should be more pronounced for a generalized domain wall, like the ones considered above for random circuits, where the average charge density on the two sides is $0 < n_L < n_R < 1$. Since such states have a finite density of particles everywhere, diffusion should hold (at long times) even far from the origin. Although for such states we are more limited in the times we can approach numerically, we indeed find a better scaling collapse with $\sqrt{t}$ both for the spin and for the entanglement profile. This is shown in the lower panels of Fig.~\ref{fig:Floquet_dw_profiles_tebd}.

\bibliography{main.bbl}

\begin{thebibliography}{55}%
\makeatletter
\providecommand \@ifxundefined [1]{%
 \@ifx{#1\undefined}
}%
\providecommand \@ifnum [1]{%
 \ifnum #1\expandafter \@firstoftwo
 \else \expandafter \@secondoftwo
 \fi
}%
\providecommand \@ifx [1]{%
 \ifx #1\expandafter \@firstoftwo
 \else \expandafter \@secondoftwo
 \fi
}%
\providecommand \natexlab [1]{#1}%
\providecommand \enquote  [1]{``#1''}%
\providecommand \bibnamefont  [1]{#1}%
\providecommand \bibfnamefont [1]{#1}%
\providecommand \citenamefont [1]{#1}%
\providecommand \href@noop [0]{\@secondoftwo}%
\providecommand \href [0]{\begingroup \@sanitize@url \@href}%
\providecommand \@href[1]{\@@startlink{#1}\@@href}%
\providecommand \@@href[1]{\endgroup#1\@@endlink}%
\providecommand \@sanitize@url [0]{\catcode `\\12\catcode `\$12\catcode
  `\&12\catcode `\#12\catcode `\^12\catcode `\_12\catcode `\%12\relax}%
\providecommand \@@startlink[1]{}%
\providecommand \@@endlink[0]{}%
\providecommand \url  [0]{\begingroup\@sanitize@url \@url }%
\providecommand \@url [1]{\endgroup\@href {#1}{\urlprefix }}%
\providecommand \urlprefix  [0]{URL }%
\providecommand \Eprint [0]{\href }%
\providecommand \doibase [0]{http://dx.doi.org/}%
\providecommand \selectlanguage [0]{\@gobble}%
\providecommand \bibinfo  [0]{\@secondoftwo}%
\providecommand \bibfield  [0]{\@secondoftwo}%
\providecommand \translation [1]{[#1]}%
\providecommand \BibitemOpen [0]{}%
\providecommand \bibitemStop [0]{}%
\providecommand \bibitemNoStop [0]{.\EOS\space}%
\providecommand \EOS [0]{\spacefactor3000\relax}%
\providecommand \BibitemShut  [1]{\csname bibitem#1\endcsname}%
\let\auto@bib@innerbib\@empty
\bibitem [{\citenamefont {Jonay}\ \emph {et~al.}(2018)\citenamefont {Jonay},
  \citenamefont {Huse},\ and\ \citenamefont {Nahum}}]{JonayNahum}%
  \BibitemOpen
  \bibfield  {author} {\bibinfo {author} {\bibfnamefont {C.}~\bibnamefont
  {Jonay}}, \bibinfo {author} {\bibfnamefont {D.~A.}\ \bibnamefont {Huse}}, \
  and\ \bibinfo {author} {\bibfnamefont {A.}~\bibnamefont {Nahum}},\
  }\href@noop {} {\  (\bibinfo {year} {2018})},\ \Eprint
  {http://arxiv.org/abs/1803.00089} {arXiv:1803.00089 [cond-mat.stat-mech]}
  \BibitemShut {NoStop}%
\bibitem [{\citenamefont {Rigol}\ \emph {et~al.}(2008)\citenamefont {Rigol},
  \citenamefont {Dunjko},\ and\ \citenamefont {Olshanii}}]{Rigol2008}%
  \BibitemOpen
  \bibfield  {author} {\bibinfo {author} {\bibfnamefont {M.}~\bibnamefont
  {Rigol}}, \bibinfo {author} {\bibfnamefont {V.}~\bibnamefont {Dunjko}}, \
  and\ \bibinfo {author} {\bibfnamefont {M.}~\bibnamefont {Olshanii}},\ }\href
  {\doibase 10.1038/nature06838} {\bibfield  {journal} {\bibinfo  {journal}
  {Nature}\ }\textbf {\bibinfo {volume} {452}},\ \bibinfo {pages} {854}
  (\bibinfo {year} {2008})}\BibitemShut {NoStop}%
\bibitem [{\citenamefont {Calabrese}\ and\ \citenamefont
  {Cardy}(2006)}]{CalabreseCardy06}%
  \BibitemOpen
  \bibfield  {author} {\bibinfo {author} {\bibfnamefont {P.}~\bibnamefont
  {Calabrese}}\ and\ \bibinfo {author} {\bibfnamefont {J.~L.}\ \bibnamefont
  {Cardy}},\ }\href {\doibase 10.1103/PhysRevLett.96.136801} {\bibfield
  {journal} {\bibinfo  {journal} {Phys. Rev. Lett.}\ }\textbf {\bibinfo
  {volume} {96}},\ \bibinfo {pages} {136801} (\bibinfo {year} {2006})},\
  \Eprint {http://arxiv.org/abs/cond-mat/0601225} {arXiv:cond-mat/0601225
  [cond-mat]} \BibitemShut {NoStop}%
\bibitem [{\citenamefont {{D'Alessio}}\ \emph {et~al.}(2016)\citenamefont
  {{D'Alessio}}, \citenamefont {{Kafri}}, \citenamefont {{Polkovnikov}},\ and\
  \citenamefont {{Rigol}}}]{ETHreviewRigol16}%
  \BibitemOpen
  \bibfield  {author} {\bibinfo {author} {\bibfnamefont {L.}~\bibnamefont
  {{D'Alessio}}}, \bibinfo {author} {\bibfnamefont {Y.}~\bibnamefont
  {{Kafri}}}, \bibinfo {author} {\bibfnamefont {A.}~\bibnamefont
  {{Polkovnikov}}}, \ and\ \bibinfo {author} {\bibfnamefont {M.}~\bibnamefont
  {{Rigol}}},\ }\href {\doibase 10.1080/00018732.2016.1198134} {\bibfield
  {journal} {\bibinfo  {journal} {Advances in Physics}\ }\textbf {\bibinfo
  {volume} {65}},\ \bibinfo {pages} {239} (\bibinfo {year} {2016})},\ \Eprint
  {http://arxiv.org/abs/1509.06411} {arXiv:1509.06411 [cond-mat.stat-mech]}
  \BibitemShut {NoStop}%
\bibitem [{\citenamefont {Gogolin}\ and\ \citenamefont
  {Eisert}(2016)}]{GogolinReview}%
  \BibitemOpen
  \bibfield  {author} {\bibinfo {author} {\bibfnamefont {C.}~\bibnamefont
  {Gogolin}}\ and\ \bibinfo {author} {\bibfnamefont {J.}~\bibnamefont
  {Eisert}},\ }\href {http://stacks.iop.org/0034-4885/79/i=5/a=056001}
  {\bibfield  {journal} {\bibinfo  {journal} {Reports on Progress in Physics}\
  }\textbf {\bibinfo {volume} {79}},\ \bibinfo {pages} {056001} (\bibinfo
  {year} {2016})}\BibitemShut {NoStop}%
\bibitem [{\citenamefont {Kaufman}\ \emph {et~al.}(2016)\citenamefont
  {Kaufman}, \citenamefont {Tai}, \citenamefont {Lukin}, \citenamefont
  {Rispoli}, \citenamefont {Schittko}, \citenamefont {Preiss},\ and\
  \citenamefont {Greiner}}]{Kaufman794}%
  \BibitemOpen
  \bibfield  {author} {\bibinfo {author} {\bibfnamefont {A.~M.}\ \bibnamefont
  {Kaufman}}, \bibinfo {author} {\bibfnamefont {M.~E.}\ \bibnamefont {Tai}},
  \bibinfo {author} {\bibfnamefont {A.}~\bibnamefont {Lukin}}, \bibinfo
  {author} {\bibfnamefont {M.}~\bibnamefont {Rispoli}}, \bibinfo {author}
  {\bibfnamefont {R.}~\bibnamefont {Schittko}}, \bibinfo {author}
  {\bibfnamefont {P.~M.}\ \bibnamefont {Preiss}}, \ and\ \bibinfo {author}
  {\bibfnamefont {M.}~\bibnamefont {Greiner}},\ }\href {\doibase
  10.1126/science.aaf6725} {\bibfield  {journal} {\bibinfo  {journal}
  {Science}\ }\textbf {\bibinfo {volume} {353}},\ \bibinfo {pages} {794}
  (\bibinfo {year} {2016})},\ \Eprint
  {http://arxiv.org/abs/http://science.sciencemag.org/content/353/6301/794.full.pdf}
  {http://science.sciencemag.org/content/353/6301/794.full.pdf} \BibitemShut
  {NoStop}%
\bibitem [{\citenamefont {Deutsch}(1991)}]{Deutsch91}%
  \BibitemOpen
  \bibfield  {author} {\bibinfo {author} {\bibfnamefont {J.~M.}\ \bibnamefont
  {Deutsch}},\ }\href {\doibase 10.1103/PhysRevA.43.2046} {\bibfield  {journal}
  {\bibinfo  {journal} {Phys. Rev. A}\ }\textbf {\bibinfo {volume} {43}},\
  \bibinfo {pages} {2046} (\bibinfo {year} {1991})}\BibitemShut {NoStop}%
\bibitem [{\citenamefont {Srednicki}(1994)}]{Srednicki94}%
  \BibitemOpen
  \bibfield  {author} {\bibinfo {author} {\bibfnamefont {M.}~\bibnamefont
  {Srednicki}},\ }\href {\doibase 10.1103/PhysRevE.50.888} {\bibfield
  {journal} {\bibinfo  {journal} {Phys. Rev. E}\ }\textbf {\bibinfo {volume}
  {50}},\ \bibinfo {pages} {888} (\bibinfo {year} {1994})}\BibitemShut
  {NoStop}%
\bibitem [{\citenamefont {Abanin}\ and\ \citenamefont
  {Demler}(2012)}]{DemlerEntanglement}%
  \BibitemOpen
  \bibfield  {author} {\bibinfo {author} {\bibfnamefont {D.~A.}\ \bibnamefont
  {Abanin}}\ and\ \bibinfo {author} {\bibfnamefont {E.}~\bibnamefont
  {Demler}},\ }\href {\doibase 10.1103/PhysRevLett.109.020504} {\bibfield
  {journal} {\bibinfo  {journal} {Phys. Rev. Lett.}\ }\textbf {\bibinfo
  {volume} {109}},\ \bibinfo {pages} {020504} (\bibinfo {year}
  {2012})}\BibitemShut {NoStop}%
\bibitem [{\citenamefont {Daley}\ \emph {et~al.}(2012)\citenamefont {Daley},
  \citenamefont {Pichler}, \citenamefont {Schachenmayer},\ and\ \citenamefont
  {Zoller}}]{ZollerEntanglement}%
  \BibitemOpen
  \bibfield  {author} {\bibinfo {author} {\bibfnamefont {A.~J.}\ \bibnamefont
  {Daley}}, \bibinfo {author} {\bibfnamefont {H.}~\bibnamefont {Pichler}},
  \bibinfo {author} {\bibfnamefont {J.}~\bibnamefont {Schachenmayer}}, \ and\
  \bibinfo {author} {\bibfnamefont {P.}~\bibnamefont {Zoller}},\ }\href
  {\doibase 10.1103/PhysRevLett.109.020505} {\bibfield  {journal} {\bibinfo
  {journal} {Phys. Rev. Lett.}\ }\textbf {\bibinfo {volume} {109}},\ \bibinfo
  {pages} {020505} (\bibinfo {year} {2012})}\BibitemShut {NoStop}%
\bibitem [{\citenamefont {Islam}\ \emph {et~al.}(2015)\citenamefont {Islam},
  \citenamefont {Ma}, \citenamefont {Preiss}, \citenamefont {Eric~Tai},
  \citenamefont {Lukin}, \citenamefont {Rispoli},\ and\ \citenamefont
  {Greiner}}]{Islam15}%
  \BibitemOpen
  \bibfield  {author} {\bibinfo {author} {\bibfnamefont {R.}~\bibnamefont
  {Islam}}, \bibinfo {author} {\bibfnamefont {R.}~\bibnamefont {Ma}}, \bibinfo
  {author} {\bibfnamefont {P.}~\bibnamefont {Preiss}}, \bibinfo {author}
  {\bibfnamefont {M.}~\bibnamefont {Eric~Tai}}, \bibinfo {author}
  {\bibfnamefont {A.}~\bibnamefont {Lukin}}, \bibinfo {author} {\bibfnamefont
  {M.}~\bibnamefont {Rispoli}}, \ and\ \bibinfo {author} {\bibfnamefont
  {M.}~\bibnamefont {Greiner}},\ }\href@noop {} {\  (\bibinfo {year}
  {2015})}\BibitemShut {NoStop}%
\bibitem [{\citenamefont {Elben}\ \emph {et~al.}(2018)\citenamefont {Elben},
  \citenamefont {Vermersch}, \citenamefont {Dalmonte}, \citenamefont {Cirac},\
  and\ \citenamefont {Zoller}}]{Elben18}%
  \BibitemOpen
  \bibfield  {author} {\bibinfo {author} {\bibfnamefont {A.}~\bibnamefont
  {Elben}}, \bibinfo {author} {\bibfnamefont {B.}~\bibnamefont {Vermersch}},
  \bibinfo {author} {\bibfnamefont {M.}~\bibnamefont {Dalmonte}}, \bibinfo
  {author} {\bibfnamefont {J.~I.}\ \bibnamefont {Cirac}}, \ and\ \bibinfo
  {author} {\bibfnamefont {P.}~\bibnamefont {Zoller}},\ }\href {\doibase
  10.1103/PhysRevLett.120.050406} {\bibfield  {journal} {\bibinfo  {journal}
  {Phys. Rev. Lett.}\ }\textbf {\bibinfo {volume} {120}},\ \bibinfo {pages}
  {050406} (\bibinfo {year} {2018})}\BibitemShut {NoStop}%
\bibitem [{\citenamefont {Calabrese}\ and\ \citenamefont
  {Cardy}(2005)}]{CalabreseCardy05}%
  \BibitemOpen
  \bibfield  {author} {\bibinfo {author} {\bibfnamefont {P.}~\bibnamefont
  {Calabrese}}\ and\ \bibinfo {author} {\bibfnamefont {J.}~\bibnamefont
  {Cardy}},\ }\href {http://stacks.iop.org/1742-5468/2005/i=04/a=P04010}
  {\bibfield  {journal} {\bibinfo  {journal} {Journal of Statistical Mechanics:
  Theory and Experiment}\ }\textbf {\bibinfo {volume} {2005}},\ \bibinfo
  {pages} {P04010} (\bibinfo {year} {2005})}\BibitemShut {NoStop}%
\bibitem [{\citenamefont {Calabrese}\ and\ \citenamefont
  {Cardy}(2007)}]{CalabreseCardy07}%
  \BibitemOpen
  \bibfield  {author} {\bibinfo {author} {\bibfnamefont {P.}~\bibnamefont
  {Calabrese}}\ and\ \bibinfo {author} {\bibfnamefont {J.}~\bibnamefont
  {Cardy}},\ }\href {http://stacks.iop.org/1742-5468/2007/i=10/a=P10004}
  {\bibfield  {journal} {\bibinfo  {journal} {Journal of Statistical Mechanics:
  Theory and Experiment}\ }\textbf {\bibinfo {volume} {2007}},\ \bibinfo
  {pages} {P10004} (\bibinfo {year} {2007})}\BibitemShut {NoStop}%
\bibitem [{\citenamefont {Alba}\ and\ \citenamefont
  {Calabrese}(2017)}]{Alba2017_1}%
  \BibitemOpen
  \bibfield  {author} {\bibinfo {author} {\bibfnamefont {V.}~\bibnamefont
  {Alba}}\ and\ \bibinfo {author} {\bibfnamefont {P.}~\bibnamefont
  {Calabrese}},\ }\href {\doibase 10.1073/pnas.1703516114} {\bibfield
  {journal} {\bibinfo  {journal} {Proceedings of the National Academy of
  Sciences}\ }\textbf {\bibinfo {volume} {114}},\ \bibinfo {pages} {7947}
  (\bibinfo {year} {2017})},\ \Eprint
  {http://arxiv.org/abs/http://www.pnas.org/content/114/30/7947.full.pdf}
  {http://www.pnas.org/content/114/30/7947.full.pdf} \BibitemShut {NoStop}%
\bibitem [{\citenamefont {Alba}\ and\ \citenamefont
  {Calabrese}(2018)}]{Alba2017_2}%
  \BibitemOpen
  \bibfield  {author} {\bibinfo {author} {\bibfnamefont {V.}~\bibnamefont
  {Alba}}\ and\ \bibinfo {author} {\bibfnamefont {P.}~\bibnamefont
  {Calabrese}},\ }\href {\doibase 10.21468/SciPostPhys.4.3.017} {\bibfield
  {journal} {\bibinfo  {journal} {SciPost Phys.}\ }\textbf {\bibinfo {volume}
  {4}},\ \bibinfo {pages} {17} (\bibinfo {year} {2018})}\BibitemShut {NoStop}%
\bibitem [{\citenamefont {Kim}\ and\ \citenamefont
  {Huse}(2013)}]{HyungwonHuse}%
  \BibitemOpen
  \bibfield  {author} {\bibinfo {author} {\bibfnamefont {H.}~\bibnamefont
  {Kim}}\ and\ \bibinfo {author} {\bibfnamefont {D.~A.}\ \bibnamefont {Huse}},\
  }\href {\doibase 10.1103/PhysRevLett.111.127205} {\bibfield  {journal}
  {\bibinfo  {journal} {Phys. Rev. Lett.}\ }\textbf {\bibinfo {volume} {111}},\
  \bibinfo {pages} {127205} (\bibinfo {year} {2013})}\BibitemShut {NoStop}%
\bibitem [{\citenamefont {Nahum}\ \emph {et~al.}(2017)\citenamefont {Nahum},
  \citenamefont {Ruhman}, \citenamefont {Vijay},\ and\ \citenamefont
  {Haah}}]{Nahum16}%
  \BibitemOpen
  \bibfield  {author} {\bibinfo {author} {\bibfnamefont {A.}~\bibnamefont
  {Nahum}}, \bibinfo {author} {\bibfnamefont {J.}~\bibnamefont {Ruhman}},
  \bibinfo {author} {\bibfnamefont {S.}~\bibnamefont {Vijay}}, \ and\ \bibinfo
  {author} {\bibfnamefont {J.}~\bibnamefont {Haah}},\ }\href {\doibase
  10.1103/PhysRevX.7.031016} {\bibfield  {journal} {\bibinfo  {journal} {Phys.
  Rev. X}\ }\textbf {\bibinfo {volume} {7}},\ \bibinfo {pages} {031016}
  (\bibinfo {year} {2017})}\BibitemShut {NoStop}%
\bibitem [{\citenamefont {Nahum}\ \emph
  {et~al.}(2018{\natexlab{a}})\citenamefont {Nahum}, \citenamefont {Vijay},\
  and\ \citenamefont {Haah}}]{Nahum17}%
  \BibitemOpen
  \bibfield  {author} {\bibinfo {author} {\bibfnamefont {A.}~\bibnamefont
  {Nahum}}, \bibinfo {author} {\bibfnamefont {S.}~\bibnamefont {Vijay}}, \ and\
  \bibinfo {author} {\bibfnamefont {J.}~\bibnamefont {Haah}},\ }\href {\doibase
  10.1103/PhysRevX.8.021014} {\bibfield  {journal} {\bibinfo  {journal} {Phys.
  Rev. X}\ }\textbf {\bibinfo {volume} {8}},\ \bibinfo {pages} {021014}
  (\bibinfo {year} {2018}{\natexlab{a}})}\BibitemShut {NoStop}%
\bibitem [{\citenamefont {Zhou}\ and\ \citenamefont {Nahum}(2018)}]{ZhouNahum}%
  \BibitemOpen
  \bibfield  {author} {\bibinfo {author} {\bibfnamefont {T.}~\bibnamefont
  {Zhou}}\ and\ \bibinfo {author} {\bibfnamefont {A.}~\bibnamefont {Nahum}},\
  }\href@noop {} {\  (\bibinfo {year} {2018})},\ \Eprint
  {http://arxiv.org/abs/1804.09737} {arXiv:1804.09737 [cond-mat.stat-mech]}
  \BibitemShut {NoStop}%
\bibitem [{\citenamefont {Mezei}(2018)}]{Mezei18}%
  \BibitemOpen
  \bibfield  {author} {\bibinfo {author} {\bibfnamefont {M.}~\bibnamefont
  {Mezei}},\ }\href {\doibase 10.1103/PhysRevD.98.106025} {\bibfield  {journal}
  {\bibinfo  {journal} {Phys. Rev. D}\ }\textbf {\bibinfo {volume} {98}},\
  \bibinfo {pages} {106025} (\bibinfo {year} {2018})}\BibitemShut {NoStop}%
\bibitem [{\citenamefont {Bertini}\ \emph
  {et~al.}(2018{\natexlab{a}})\citenamefont {Bertini}, \citenamefont {Kos},\
  and\ \citenamefont {Prosen}}]{Bertini2018}%
  \BibitemOpen
  \bibfield  {author} {\bibinfo {author} {\bibfnamefont {B.}~\bibnamefont
  {Bertini}}, \bibinfo {author} {\bibfnamefont {P.}~\bibnamefont {Kos}}, \ and\
  \bibinfo {author} {\bibfnamefont {T.}~\bibnamefont {Prosen}},\ }\href@noop {}
  {\  (\bibinfo {year} {2018}{\natexlab{a}})},\ \Eprint
  {http://arxiv.org/abs/1812.05090} {arXiv:1812.05090 [cond-mat.stat-mech]}
  \BibitemShut {NoStop}%
\bibitem [{\citenamefont {Rakovszky}\ \emph {et~al.}(2019)\citenamefont
  {Rakovszky}, \citenamefont {Pollmann},\ and\ \citenamefont {von
  Keyserlingk}}]{DiffusiveRenyi}%
  \BibitemOpen
  \bibfield  {author} {\bibinfo {author} {\bibfnamefont {T.}~\bibnamefont
  {Rakovszky}}, \bibinfo {author} {\bibfnamefont {F.}~\bibnamefont {Pollmann}},
  \ and\ \bibinfo {author} {\bibfnamefont {C.~W.}\ \bibnamefont {von
  Keyserlingk}},\ }\href@noop {} {\enquote {\bibinfo {title} {Sub-ballistic
  growth of r\'enyi entropies due to diffusion},}\ } (\bibinfo {year} {2019}),\
  \Eprint {http://arxiv.org/abs/arXiv:1901.10502} {arXiv:1901.10502}
  \BibitemShut {NoStop}%
\bibitem [{\citenamefont {Eisler}\ \emph {et~al.}(2009)\citenamefont {Eisler},
  \citenamefont {Igl\'oi},\ and\ \citenamefont {Peschel}}]{Eisler2009}%
  \BibitemOpen
  \bibfield  {author} {\bibinfo {author} {\bibfnamefont {V.}~\bibnamefont
  {Eisler}}, \bibinfo {author} {\bibfnamefont {F.}~\bibnamefont {Igl\'oi}}, \
  and\ \bibinfo {author} {\bibfnamefont {I.}~\bibnamefont {Peschel}},\ }\href
  {http://stacks.iop.org/1742-5468/2009/i=02/a=P02011} {\bibfield  {journal}
  {\bibinfo  {journal} {Journal of Statistical Mechanics: Theory and
  Experiment}\ }\textbf {\bibinfo {volume} {2009}},\ \bibinfo {pages} {P02011}
  (\bibinfo {year} {2009})}\BibitemShut {NoStop}%
\bibitem [{\citenamefont {Alba}\ and\ \citenamefont
  {Heidrich-Meisner}(2014)}]{Alba2014}%
  \BibitemOpen
  \bibfield  {author} {\bibinfo {author} {\bibfnamefont {V.}~\bibnamefont
  {Alba}}\ and\ \bibinfo {author} {\bibfnamefont {F.}~\bibnamefont
  {Heidrich-Meisner}},\ }\href {\doibase 10.1103/PhysRevB.90.075144} {\bibfield
   {journal} {\bibinfo  {journal} {Phys. Rev. B}\ }\textbf {\bibinfo {volume}
  {90}},\ \bibinfo {pages} {075144} (\bibinfo {year} {2014})}\BibitemShut
  {NoStop}%
\bibitem [{\citenamefont {Vidmar}\ \emph {et~al.}(2017)\citenamefont {Vidmar},
  \citenamefont {Iyer},\ and\ \citenamefont {Rigol}}]{Vidmar2017}%
  \BibitemOpen
  \bibfield  {author} {\bibinfo {author} {\bibfnamefont {L.}~\bibnamefont
  {Vidmar}}, \bibinfo {author} {\bibfnamefont {D.}~\bibnamefont {Iyer}}, \ and\
  \bibinfo {author} {\bibfnamefont {M.}~\bibnamefont {Rigol}},\ }\href
  {\doibase 10.1103/PhysRevX.7.021012} {\bibfield  {journal} {\bibinfo
  {journal} {Phys. Rev. X}\ }\textbf {\bibinfo {volume} {7}},\ \bibinfo {pages}
  {021012} (\bibinfo {year} {2017})}\BibitemShut {NoStop}%
\bibitem [{\citenamefont {Ljubotina}\ \emph {et~al.}(2017)\citenamefont
  {Ljubotina}, \citenamefont {Znidaric},\ and\ \citenamefont
  {Prosen}}]{Ljubotina2017}%
  \BibitemOpen
  \bibfield  {author} {\bibinfo {author} {\bibfnamefont {M.}~\bibnamefont
  {Ljubotina}}, \bibinfo {author} {\bibfnamefont {M.}~\bibnamefont {Znidaric}},
  \ and\ \bibinfo {author} {\bibfnamefont {T.}~\bibnamefont {Prosen}},\ }\href
  {\doibase 10.1038/ncomms16117} {\bibfield  {journal} {\bibinfo  {journal}
  {Nature Communications}\ }\textbf {\bibinfo {volume} {8}} (\bibinfo {year}
  {2017}),\ 10.1038/ncomms16117}\BibitemShut {NoStop}%
\bibitem [{\citenamefont {B.~Bulchandani}\ and\ \citenamefont
  {Karrasch}(2018)}]{Karrasch2018}%
  \BibitemOpen
  \bibfield  {author} {\bibinfo {author} {\bibfnamefont {V.}~\bibnamefont
  {B.~Bulchandani}}\ and\ \bibinfo {author} {\bibfnamefont {C.}~\bibnamefont
  {Karrasch}},\ }\href@noop {} {\bibfield  {journal} {\bibinfo  {journal}
  {arXiv preprint arXiv:1810.08227}\ } (\bibinfo {year} {2018})}\BibitemShut
  {NoStop}%
\bibitem [{\citenamefont {Bertini}\ \emph
  {et~al.}(2018{\natexlab{b}})\citenamefont {Bertini}, \citenamefont {Fagotti},
  \citenamefont {Piroli},\ and\ \citenamefont {Calabrese}}]{Bertini_2018}%
  \BibitemOpen
  \bibfield  {author} {\bibinfo {author} {\bibfnamefont {B.}~\bibnamefont
  {Bertini}}, \bibinfo {author} {\bibfnamefont {M.}~\bibnamefont {Fagotti}},
  \bibinfo {author} {\bibfnamefont {L.}~\bibnamefont {Piroli}}, \ and\ \bibinfo
  {author} {\bibfnamefont {P.}~\bibnamefont {Calabrese}},\ }\href {\doibase
  10.1088/1751-8121/aad82e} {\bibfield  {journal} {\bibinfo  {journal} {Journal
  of Physics A: Mathematical and Theoretical}\ }\textbf {\bibinfo {volume}
  {51}},\ \bibinfo {pages} {39LT01} (\bibinfo {year}
  {2018}{\natexlab{b}})}\BibitemShut {NoStop}%
\bibitem [{\citenamefont {Alba}\ \emph {et~al.}(2019)\citenamefont {Alba},
  \citenamefont {Bertini},\ and\ \citenamefont {Fagotti}}]{Bertini_2019}%
  \BibitemOpen
  \bibfield  {author} {\bibinfo {author} {\bibfnamefont {V.}~\bibnamefont
  {Alba}}, \bibinfo {author} {\bibfnamefont {B.}~\bibnamefont {Bertini}}, \
  and\ \bibinfo {author} {\bibfnamefont {M.}~\bibnamefont {Fagotti}},\
  }\href@noop {} {\enquote {\bibinfo {title} {Entanglement evolution and
  generalised hydrodynamics: interacting integrable systems},}\ } (\bibinfo
  {year} {2019}),\ \Eprint {http://arxiv.org/abs/arXiv:1903.00467}
  {arXiv:1903.00467} \BibitemShut {NoStop}%
\bibitem [{\citenamefont {{Mesty{\'a}n}}\ and\ \citenamefont
  {{Alba}}(2019)}]{Mestyan_2019}%
  \BibitemOpen
  \bibfield  {author} {\bibinfo {author} {\bibfnamefont {M.}~\bibnamefont
  {{Mesty{\'a}n}}}\ and\ \bibinfo {author} {\bibfnamefont {V.}~\bibnamefont
  {{Alba}}},\ }\href@noop {} {\enquote {\bibinfo {title} {Molecular dynamics
  simulation of entanglement spreading in generalized hydrodynamics},}\ }
  (\bibinfo {year} {2019}),\ \Eprint {http://arxiv.org/abs/arXiv:1905.03206}
  {arXiv:1905.03206} \BibitemShut {NoStop}%
\bibitem [{\citenamefont {Bloembergen}(1949)}]{BLOEMBERGEN1949}%
  \BibitemOpen
  \bibfield  {author} {\bibinfo {author} {\bibfnamefont {N.}~\bibnamefont
  {Bloembergen}},\ }\href {\doibase
  https://doi.org/10.1016/0031-8914(49)90114-7} {\bibfield  {journal} {\bibinfo
   {journal} {Physica}\ }\textbf {\bibinfo {volume} {15}},\ \bibinfo {pages}
  {386 } (\bibinfo {year} {1949})}\BibitemShut {NoStop}%
\bibitem [{\citenamefont {Gennes}(1958)}]{DEGENNES1958}%
  \BibitemOpen
  \bibfield  {author} {\bibinfo {author} {\bibfnamefont {P.~D.}\ \bibnamefont
  {Gennes}},\ }\href {\doibase https://doi.org/10.1016/0022-3697(58)90120-3}
  {\bibfield  {journal} {\bibinfo  {journal} {Journal of Physics and Chemistry
  of Solids}\ }\textbf {\bibinfo {volume} {4}},\ \bibinfo {pages} {223 }
  (\bibinfo {year} {1958})}\BibitemShut {NoStop}%
\bibitem [{\citenamefont {Kadanoff}\ and\ \citenamefont
  {Martin}(1963)}]{KADANOFF1963}%
  \BibitemOpen
  \bibfield  {author} {\bibinfo {author} {\bibfnamefont {L.~P.}\ \bibnamefont
  {Kadanoff}}\ and\ \bibinfo {author} {\bibfnamefont {P.~C.}\ \bibnamefont
  {Martin}},\ }\href {\doibase https://doi.org/10.1016/0003-4916(63)90078-2}
  {\bibfield  {journal} {\bibinfo  {journal} {Annals of Physics}\ }\textbf
  {\bibinfo {volume} {24}},\ \bibinfo {pages} {419 } (\bibinfo {year}
  {1963})}\BibitemShut {NoStop}%
\bibitem [{\citenamefont {Bohrdt}\ \emph {et~al.}(2017)\citenamefont {Bohrdt},
  \citenamefont {Mendl}, \citenamefont {Endres},\ and\ \citenamefont
  {Knap}}]{Bohrdt16}%
  \BibitemOpen
  \bibfield  {author} {\bibinfo {author} {\bibfnamefont {A.}~\bibnamefont
  {Bohrdt}}, \bibinfo {author} {\bibfnamefont {C.~B.}\ \bibnamefont {Mendl}},
  \bibinfo {author} {\bibfnamefont {M.}~\bibnamefont {Endres}}, \ and\ \bibinfo
  {author} {\bibfnamefont {M.}~\bibnamefont {Knap}},\ }\href
  {http://stacks.iop.org/1367-2630/19/i=6/a=063001} {\bibfield  {journal}
  {\bibinfo  {journal} {New Journal of Physics}\ }\textbf {\bibinfo {volume}
  {19}},\ \bibinfo {pages} {063001} (\bibinfo {year} {2017})}\BibitemShut
  {NoStop}%
\bibitem [{\citenamefont {Khemani}\ \emph {et~al.}(2018)\citenamefont
  {Khemani}, \citenamefont {Vishwanath},\ and\ \citenamefont
  {Huse}}]{OTOCDiff1}%
  \BibitemOpen
  \bibfield  {author} {\bibinfo {author} {\bibfnamefont {V.}~\bibnamefont
  {Khemani}}, \bibinfo {author} {\bibfnamefont {A.}~\bibnamefont {Vishwanath}},
  \ and\ \bibinfo {author} {\bibfnamefont {D.~A.}\ \bibnamefont {Huse}},\
  }\href {\doibase 10.1103/PhysRevX.8.031057} {\bibfield  {journal} {\bibinfo
  {journal} {Phys. Rev. X}\ }\textbf {\bibinfo {volume} {8}},\ \bibinfo {pages}
  {031057} (\bibinfo {year} {2018})}\BibitemShut {NoStop}%
\bibitem [{\citenamefont {Rakovszky}\ \emph {et~al.}(2018)\citenamefont
  {Rakovszky}, \citenamefont {Pollmann},\ and\ \citenamefont {von
  Keyserlingk}}]{OTOCDiff2}%
  \BibitemOpen
  \bibfield  {author} {\bibinfo {author} {\bibfnamefont {T.}~\bibnamefont
  {Rakovszky}}, \bibinfo {author} {\bibfnamefont {F.}~\bibnamefont {Pollmann}},
  \ and\ \bibinfo {author} {\bibfnamefont {C.~W.}\ \bibnamefont {von
  Keyserlingk}},\ }\href {\doibase 10.1103/PhysRevX.8.031058} {\bibfield
  {journal} {\bibinfo  {journal} {Phys. Rev. X}\ }\textbf {\bibinfo {volume}
  {8}},\ \bibinfo {pages} {031058} (\bibinfo {year} {2018})}\BibitemShut
  {NoStop}%
\bibitem [{\citenamefont {von Keyserlingk}\ \emph {et~al.}(2018)\citenamefont
  {von Keyserlingk}, \citenamefont {Rakovszky}, \citenamefont {Pollmann},\ and\
  \citenamefont {Sondhi}}]{RvK17}%
  \BibitemOpen
  \bibfield  {author} {\bibinfo {author} {\bibfnamefont {C.~W.}\ \bibnamefont
  {von Keyserlingk}}, \bibinfo {author} {\bibfnamefont {T.}~\bibnamefont
  {Rakovszky}}, \bibinfo {author} {\bibfnamefont {F.}~\bibnamefont {Pollmann}},
  \ and\ \bibinfo {author} {\bibfnamefont {S.~L.}\ \bibnamefont {Sondhi}},\
  }\href {\doibase 10.1103/PhysRevX.8.021013} {\bibfield  {journal} {\bibinfo
  {journal} {Phys. Rev. X}\ }\textbf {\bibinfo {volume} {8}},\ \bibinfo {pages}
  {021013} (\bibinfo {year} {2018})}\BibitemShut {NoStop}%
\bibitem [{\citenamefont {Kardar}\ \emph {et~al.}(1986)\citenamefont {Kardar},
  \citenamefont {Parisi},\ and\ \citenamefont {Zhang}}]{KPZ}%
  \BibitemOpen
  \bibfield  {author} {\bibinfo {author} {\bibfnamefont {M.}~\bibnamefont
  {Kardar}}, \bibinfo {author} {\bibfnamefont {G.}~\bibnamefont {Parisi}}, \
  and\ \bibinfo {author} {\bibfnamefont {Y.-C.}\ \bibnamefont {Zhang}},\ }\href
  {\doibase 10.1103/PhysRevLett.56.889} {\bibfield  {journal} {\bibinfo
  {journal} {Phys. Rev. Lett.}\ }\textbf {\bibinfo {volume} {56}},\ \bibinfo
  {pages} {889} (\bibinfo {year} {1986})}\BibitemShut {NoStop}%
\bibitem [{\citenamefont {Lukin}\ \emph {et~al.}(2018)\citenamefont {Lukin},
  \citenamefont {Rispoli}, \citenamefont {Schittko}, \citenamefont {Eric~Tai},
  \citenamefont {M.~Kaufman}, \citenamefont {Choi}, \citenamefont {Khemani},
  \citenamefont {Leonard},\ and\ \citenamefont {Greiner}}]{Lukin18}%
  \BibitemOpen
  \bibfield  {author} {\bibinfo {author} {\bibfnamefont {A.}~\bibnamefont
  {Lukin}}, \bibinfo {author} {\bibfnamefont {M.}~\bibnamefont {Rispoli}},
  \bibinfo {author} {\bibfnamefont {R.}~\bibnamefont {Schittko}}, \bibinfo
  {author} {\bibfnamefont {M.}~\bibnamefont {Eric~Tai}}, \bibinfo {author}
  {\bibfnamefont {A.}~\bibnamefont {M.~Kaufman}}, \bibinfo {author}
  {\bibfnamefont {S.}~\bibnamefont {Choi}}, \bibinfo {author} {\bibfnamefont
  {V.}~\bibnamefont {Khemani}}, \bibinfo {author} {\bibfnamefont
  {J.}~\bibnamefont {Leonard}}, \ and\ \bibinfo {author} {\bibfnamefont
  {M.}~\bibnamefont {Greiner}},\ }\href@noop {} {\bibfield  {journal} {\bibinfo
   {journal} {arXiv preprint arXiv:1506.00650}\ } (\bibinfo {year}
  {2018})}\BibitemShut {NoStop}%
\bibitem [{\citenamefont {Nahum}\ \emph
  {et~al.}(2018{\natexlab{b}})\citenamefont {Nahum}, \citenamefont {Ruhman},\
  and\ \citenamefont {Huse}}]{Nahum18}%
  \BibitemOpen
  \bibfield  {author} {\bibinfo {author} {\bibfnamefont {A.}~\bibnamefont
  {Nahum}}, \bibinfo {author} {\bibfnamefont {J.}~\bibnamefont {Ruhman}}, \
  and\ \bibinfo {author} {\bibfnamefont {D.~A.}\ \bibnamefont {Huse}},\ }\href
  {\doibase 10.1103/PhysRevB.98.035118} {\bibfield  {journal} {\bibinfo
  {journal} {Phys. Rev. B}\ }\textbf {\bibinfo {volume} {98}},\ \bibinfo
  {pages} {035118} (\bibinfo {year} {2018}{\natexlab{b}})}\BibitemShut
  {NoStop}%
\bibitem [{\citenamefont {Chan}\ \emph
  {et~al.}(2018{\natexlab{a}})\citenamefont {Chan}, \citenamefont {De~Luca},\
  and\ \citenamefont {Chalker}}]{ChanDeLuca1}%
  \BibitemOpen
  \bibfield  {author} {\bibinfo {author} {\bibfnamefont {A.}~\bibnamefont
  {Chan}}, \bibinfo {author} {\bibfnamefont {A.}~\bibnamefont {De~Luca}}, \
  and\ \bibinfo {author} {\bibfnamefont {J.~T.}\ \bibnamefont {Chalker}},\
  }\href {\doibase 10.1103/PhysRevX.8.041019} {\bibfield  {journal} {\bibinfo
  {journal} {Phys. Rev. X}\ }\textbf {\bibinfo {volume} {8}},\ \bibinfo {pages}
  {041019} (\bibinfo {year} {2018}{\natexlab{a}})}\BibitemShut {NoStop}%
\bibitem [{\citenamefont {Chan}\ \emph
  {et~al.}(2018{\natexlab{b}})\citenamefont {Chan}, \citenamefont {De~Luca},\
  and\ \citenamefont {Chalker}}]{ChanDeLuca2}%
  \BibitemOpen
  \bibfield  {author} {\bibinfo {author} {\bibfnamefont {A.}~\bibnamefont
  {Chan}}, \bibinfo {author} {\bibfnamefont {A.}~\bibnamefont {De~Luca}}, \
  and\ \bibinfo {author} {\bibfnamefont {J.~T.}\ \bibnamefont {Chalker}},\
  }\href {\doibase 10.1103/PhysRevLett.121.060601} {\bibfield  {journal}
  {\bibinfo  {journal} {Phys. Rev. Lett.}\ }\textbf {\bibinfo {volume} {121}},\
  \bibinfo {pages} {060601} (\bibinfo {year} {2018}{\natexlab{b}})}\BibitemShut
  {NoStop}%
\bibitem [{\citenamefont {Chan}\ \emph {et~al.}(2019)\citenamefont {Chan},
  \citenamefont {De~Luca},\ and\ \citenamefont {Chalker}}]{ChanDeLuca3}%
  \BibitemOpen
  \bibfield  {author} {\bibinfo {author} {\bibfnamefont {A.}~\bibnamefont
  {Chan}}, \bibinfo {author} {\bibfnamefont {A.}~\bibnamefont {De~Luca}}, \
  and\ \bibinfo {author} {\bibfnamefont {J.~T.}\ \bibnamefont {Chalker}},\
  }\href {\doibase 10.1103/PhysRevLett.122.220601} {\bibfield  {journal}
  {\bibinfo  {journal} {Phys. Rev. Lett.}\ }\textbf {\bibinfo {volume} {122}},\
  \bibinfo {pages} {220601} (\bibinfo {year} {2019})}\BibitemShut {NoStop}%
\bibitem [{Note1()}]{Note1}%
  \BibitemOpen
  \bibinfo {note} {Note that this is in contrast with R\'enyi entropies
  $S_{\alpha > 1}$, which are strongly influenced by \protect \emph
  {fluctuations} of the conserved charge, even for homogenous quenches, and can
  grow sub-ballistically as a consequence~\cite {DiffusiveRenyi}}\BibitemShut
  {NoStop}%
\bibitem [{Note2()}]{Note2}%
  \BibitemOpen
  \bibinfo {note} {While this parameter counting argument is not rigorous, in
  the sense that fine-tuned states or unitaries could violate it, it is
  expected to hold for generic states, which we indeed find numerically for
  random MPS.}\BibitemShut {Stop}%
\bibitem [{\citenamefont {T{\"a}uber}(2007)}]{Tauber2007}%
  \BibitemOpen
  \bibfield  {author} {\bibinfo {author} {\bibfnamefont {U.~C.}\ \bibnamefont
  {T{\"a}uber}},\ }\enquote {\bibinfo {title} {Field-theory approaches to
  nonequilibrium dynamics},}\ in\ \href {\doibase 10.1007/3-540-69684-9_7}
  {\emph {\bibinfo {booktitle} {Ageing and the Glass Transition}}},\ \bibinfo
  {editor} {edited by\ \bibinfo {editor} {\bibfnamefont {M.}~\bibnamefont
  {Henkel}}, \bibinfo {editor} {\bibfnamefont {M.}~\bibnamefont {Pleimling}}, \
  and\ \bibinfo {editor} {\bibfnamefont {R.}~\bibnamefont {Sanctuary}}}\
  (\bibinfo  {publisher} {Springer Berlin Heidelberg},\ \bibinfo {address}
  {Berlin, Heidelberg},\ \bibinfo {year} {2007})\ pp.\ \bibinfo {pages}
  {295--348}\BibitemShut {NoStop}%
\bibitem [{Note3()}]{Note3}%
  \BibitemOpen
  \bibinfo {note} {This is similar to the mechanism proposed for entanglement
  growth in disordered Griffith phases where the bottleneck is provided by
  localized regions which act as `weak links', see Ref. \protect
  \rev@citealpnum {Nahum18}}\BibitemShut {NoStop}%
\bibitem [{Note4()}]{Note4}%
  \BibitemOpen
  \bibinfo {note} {This is most easily seen by taking the spatial derivative of
  the diffusion equation, which results in the same equation for $\partial _x
  \protect \mathfrak {q}$. The initial condition for $\partial _x \protect
  \mathfrak {q}$ is given by a delta function, $\partial _x \protect \mathfrak
  {q}(x,0) = \delta (x)$, for the domain wall, which becomes a Gaussian of
  width $\protect \sqrt {Dt}$ at time $t$. Integrating it up gives the result
  for $\protect \mathfrak {q}(x,t)$ stated in the text.}\BibitemShut {Stop}%
\bibitem [{\citenamefont {{Lux}}\ \emph {et~al.}(2014)\citenamefont {{Lux}},
  \citenamefont {{M{\"u}ller}}, \citenamefont {{Mitra}},\ and\ \citenamefont
  {{Rosch}}}]{Rosch13}%
  \BibitemOpen
  \bibfield  {author} {\bibinfo {author} {\bibfnamefont {J.}~\bibnamefont
  {{Lux}}}, \bibinfo {author} {\bibfnamefont {J.}~\bibnamefont {{M{\"u}ller}}},
  \bibinfo {author} {\bibfnamefont {A.}~\bibnamefont {{Mitra}}}, \ and\
  \bibinfo {author} {\bibfnamefont {A.}~\bibnamefont {{Rosch}}},\ }\href
  {\doibase 10.1103/PhysRevA.89.053608} {\bibfield  {journal} {\bibinfo
  {journal} {\pra}\ }\textbf {\bibinfo {volume} {89}},\ \bibinfo {eid} {053608}
  (\bibinfo {year} {2014})},\ \Eprint {http://arxiv.org/abs/1311.7644}
  {arXiv:1311.7644 [cond-mat.quant-gas]} \BibitemShut {NoStop}%
\bibitem [{Note5()}]{Note5}%
  \BibitemOpen
  \bibinfo {note} {Such is the case of integrable systems, which can exhibit
  logarithmic growth of the entanglement~\cite {Alba2014,Vidmar2017}, even
  though their transport properties are typically \protect \emph {faster} than
  diffusive.}\BibitemShut {Stop}%
\bibitem [{Note6()}]{Note6}%
  \BibitemOpen
  \bibinfo {note} {Note that even in the random circuit results at $N=1$, shown
  in Fig.~\ref {fig:dw_N=1}, the $~\protect \sqrt {t}$ growth is only true
  asymptotically and at short times a slightly faster growth is
  observed.}\BibitemShut {Stop}%
\bibitem [{\citenamefont {Vidal}(2003)}]{VidalTEBD}%
  \BibitemOpen
  \bibfield  {author} {\bibinfo {author} {\bibfnamefont {G.}~\bibnamefont
  {Vidal}},\ }\href {\doibase 10.1103/PhysRevLett.91.147902} {\bibfield
  {journal} {\bibinfo  {journal} {Phys. Rev. Lett.}\ }\textbf {\bibinfo
  {volume} {91}},\ \bibinfo {pages} {147902} (\bibinfo {year}
  {2003})}\BibitemShut {NoStop}%
\bibitem [{\citenamefont {Hastings}\ \emph {et~al.}(2010)\citenamefont
  {Hastings}, \citenamefont {Gonz\'alez}, \citenamefont {Kallin},\ and\
  \citenamefont {Melko}}]{HastingsSwap}%
  \BibitemOpen
  \bibfield  {author} {\bibinfo {author} {\bibfnamefont {M.~B.}\ \bibnamefont
  {Hastings}}, \bibinfo {author} {\bibfnamefont {I.}~\bibnamefont
  {Gonz\'alez}}, \bibinfo {author} {\bibfnamefont {A.~B.}\ \bibnamefont
  {Kallin}}, \ and\ \bibinfo {author} {\bibfnamefont {R.~G.}\ \bibnamefont
  {Melko}},\ }\href {\doibase 10.1103/PhysRevLett.104.157201} {\bibfield
  {journal} {\bibinfo  {journal} {Phys. Rev. Lett.}\ }\textbf {\bibinfo
  {volume} {104}},\ \bibinfo {pages} {157201} (\bibinfo {year}
  {2010})}\BibitemShut {NoStop}%
\bibitem [{\citenamefont {Hayden}\ \emph {et~al.}(2016)\citenamefont {Hayden},
  \citenamefont {Nezami}, \citenamefont {Qi}, \citenamefont {Thomas},
  \citenamefont {Walter},\ and\ \citenamefont {Yang}}]{Hayden2016}%
  \BibitemOpen
  \bibfield  {author} {\bibinfo {author} {\bibfnamefont {P.}~\bibnamefont
  {Hayden}}, \bibinfo {author} {\bibfnamefont {S.}~\bibnamefont {Nezami}},
  \bibinfo {author} {\bibfnamefont {X.-L.}\ \bibnamefont {Qi}}, \bibinfo
  {author} {\bibfnamefont {N.}~\bibnamefont {Thomas}}, \bibinfo {author}
  {\bibfnamefont {M.}~\bibnamefont {Walter}}, \ and\ \bibinfo {author}
  {\bibfnamefont {Z.}~\bibnamefont {Yang}},\ }\href {\doibase
  10.1007/JHEP11(2016)009} {\bibfield  {journal} {\bibinfo  {journal} {Journal
  of High Energy Physics}\ }\textbf {\bibinfo {volume} {2016}},\ \bibinfo
  {pages} {9} (\bibinfo {year} {2016})}\BibitemShut {NoStop}%
\end{thebibliography}%


\end{document}